\documentclass[superscriptaddress,aps,preprintnumbers,amsmath,amssymb,prd,nofootinbib,preprint]{revtex4-1}
\pdfoutput=1
\usepackage[utf8]{inputenc}
\usepackage{amsmath}
\usepackage{amsfonts}
\usepackage{amssymb}
\usepackage{graphicx}
\usepackage{physics}
\usepackage{footmisc}
\usepackage{feynmp-auto}
\usepackage{float}
\usepackage{slashed}
\usepackage[left=1 in, right=1 in, top=1 in, bottom = 1 in]{geometry}
\usepackage{hyperref}
\usepackage[dvipsnames]{xcolor}
\usepackage{comment}
\hypersetup{
    colorlinks=true,
    linkcolor=blue,
    filecolor=magenta,      	
    urlcolor=cyan,
}

\definecolor{zachcolor}{RGB}{46,85,124}

\newcommand{\eps}{\epsilon}

\begin{document}

\title{Indirect Detection of Secluded Supersymmetric Dark Matter}

\author{Patrick Barnes}
\affiliation{Leinweber Center for Theoretical Physics, Department of Physics,  University of Michigan, Ann Arbor, MI 48109, USA}
\author{Zachary Johnson}
\affiliation{Leinweber Center for Theoretical Physics, Department of Physics,  University of Michigan, Ann Arbor, MI 48109, USA}
\author{Aaron Pierce}
\affiliation{Leinweber Center for Theoretical Physics, Department of Physics,  University of Michigan, Ann Arbor, MI 48109, USA}
\author{Bibhushan Shakya}
\affiliation{Deutsches Elektronen-Synchrotron DESY, Notkestr.\,85, 22607 Hamburg, Germany}
\affiliation {CERN, Theoretical Physics Department, 1211 Geneva 23, Switzerland}
\date{\today}

\begin{abstract}
   Weak-scale secluded sector dark matter can reproduce the observed dark matter relic density with thermal freeze-out within that sector.  If nature is supersymmetric, three portals to the visible sector \textemdash ~a gauge portal, a Higgs portal, and a gaugino portal \textemdash ~are present.  We present gamma ray spectra relevant for indirect detection of dark matter annihilation in such setups. Since symmetries in the secluded sector can stabilize dark matter, $R$-parity is unnecessary, and we investigate the impact of $R$-parity violation on annihilation spectra. We present limits from the Fermi Large Area Telescope observations of dwarf galaxies and projections for Cherenkov Telescope Array observations of the galactic center.  Many of our results are also applicable to generic, non-supersymmetric setups.
\end{abstract}

\preprint{CERN-TH-2021-089}

\preprint{DESY 21-091}

\preprint{LCTP-21-13}

\maketitle 

\tableofcontents

%%%%%%%%%%%%%%%%%%%%%%%%%%%%%%%%%%%%
\noindent

\section{Introduction}
The weakly interacting massive particle (WIMP) paradigm \textemdash~wherein the relic density of the dark matter (DM) is explained by the thermal freeze-out of dark matter from the thermal bath via a weak scale annihilation cross section~\cite{Lee:1977ua} \textemdash ~remains an attractive mechanism to explain the observed abundance of dark matter in the Universe.  While the absence of signals at direct detection experiments strongly constrains traditional WIMP candidates such as the lightest supersymmetric particle (LSP)~\cite{Goldberg:1983nd} of the Minimal Supersymmetric Standard Model (MSSM), the WIMP may be realized in a secluded sector~\cite{Pospelov:2007mp}. In such scenarios, WIMP dark matter may have weak scale interactions within the secluded sector, facilitating thermal freeze-out in that sector, but couplings to the Standard Model (SM) can be limited to so-called portal couplings.  

Indirect detection is a particularly robust probe of WIMP dark matter that resides in a secluded sector. Whereas direct detection signals are suppressed by the (potentially tiny) portal couplings, dark matter annihilations proceed with weak scale cross sections, as this  sets the thermal relic density. Annihilations into secluded sector particles give visible signals via subsequent cascade decays to SM states. In this paper, we focus on the production of high energy gamma rays.  Such signals are of interest for several upcoming experiments, such as the Cherenkov Telescope Array (CTA)~\cite{Acharyya:2020sbj} and the Large High Altitude Air Shower Observatory (LHAASO)\cite{He:2020ipy}, which recently started taking data. Phenomenological investigations of secluded WIMP DM annihilation have been undertaken in several studies in the literature in various simplified setups, see, e.g., Ref.~\cite{Evans:2017kti,Hooper:2019xss,Profumo:2017obk}.
Such phenomenological studies can be sharpened within specific theoretically motivated models of secluded WIMP DM. 

If WIMP dark matter is part of a weak scale secluded sector, with only small portal couplings to the SM sector, an important question is how that sector knows about the weak scale. 
 What causes the particles in that sector to have masses comparable to those in our own?  One attractive possibility for the origin of this common mass scale is that supersymmetry (SUSY) breaking is mediated to the two sectors in a similar manner, e.g. through gravity mediation.  Then, both sectors should be near the weak scale, but with ${\mathcal O}(1)$ differences in masses, depending on details of the mediation mechanism. Models of dark matter motivated by such considerations were recently explored in Ref.~\cite{Barnes:2020vsc}.  

In such frameworks, the supersymmetric kinetic mixing between the chiral field strength superfields of the two sectors~\cite{Dienes:1996zr} gives rise to portal mixings between gauge bosons, gauginos, and Higgs bosons. The gaugino portal can have implications for dark matter phenomenology~\cite{Barnes:2020vsc},  collider physics and cosmology~\cite{Arvanitaki:2009hb,Baryakhtar:2012rz}, as well as models of baryogenesis~\cite{Pierce:2019ozl}. The existence of such portal interactions can lead to novel spectra of final states relevant for indirect detection; this will be the focus of this paper.  

In minimal supersymmetric frameworks, the stability of the LSP is necessary to provide dark matter, and $R$-parity conservation is considered the most natural possibility. But with secluded dark matter, additional particles and symmetries in the secluded sector may  provide a stable dark matter candidate, and the possibility that the LSP can decay through $R$-parity violating (RPV) operators becomes more natural (for a review of $R$ parity violation, see~\cite{Barbier:2004ez}). This allows for new dark matter annihilation spectra not normally considered in the simplest supersymmetric or secluded dark matter models.  While RPV has been considered in the context of decaying dark matter (see, e.g., ~\cite{Berezinsky:1991sp,Baltz:1997gd,Baltz:1997ar,Barbier:2004ez}), the presence of a secluded sector opens the possibility that it may be important for annihilating dark matter as well.  $R$-parity violation also allows connections with novel mechanisms of baryogenesis in this setup~\cite{Pierce:2019ozl}.  

In this paper, we discuss the indirect detection phenomenology of secluded dark matter, with particular emphasis on understanding the consequences of a supersymmetric realization, including the possibility of $R$-parity violation. We display annihilation spectra for different final states and present constraints arising from  Fermi-LAT observations of dwarf galaxies~\cite{Ackermann:2015zua}. The primary results of this paper are projections for CTA, which is particularly suited to probe this class of dark matter models. While our studies are conducted with a supersymmetric setup in mind, many results are applicable to a broader variety of secluded dark matter scenarios.

\section{Preliminaries}\label{sec:dmsignal}

In this section, we set the stage for our analysis. Our ultimate goal is to derive the reach of CTA and Fermi-LAT for annihilation of secluded dark matter. For CTA, a promising target for DM searches will be the galactic center (GC), which can be observed in the southern hemisphere by CTA-South. For Fermi-LAT, we place bounds using six years of Fermi-LAT data from 15 dwarf spheroidal satellite galaxies. 

The flux of gamma rays from annihilating dark matter (DM) is calculated as
\begin{equation}
\frac{d\Phi_{\gamma}}{dE} = \frac{\langle \sigma v \rangle}{16 \pi m_{DM}^2} \frac{dN_{\gamma}}{dE}J(\Delta \Omega).\label{eq:flux}
\end{equation}
 Here, 
$m_{DM}$ is the dark matter mass.
In this paper, we assume that DM is not self-conjugate (i.e. is not its own antiparticle); for self-conjugate dark matter, the $16$ in the denominator would be replaced with an $8$.  
In Eq.~(\ref{eq:flux}), the dependence on the particle physics of dark matter annihilation is encapsulated in two ingredients: the annihilation cross section $\langle \sigma v \rangle$, which determines the overall normalization of the signal, and the spectrum per annihilation event, $dN_{\gamma}/dE$. In our analysis, we compute $dN_{\gamma}/dE$ using the MadGraph~\cite{Alwall:2014hca} plug-in MadDM~\cite{Ambrogi:2018jqj}, with Pythia 8~\cite{Sjostrand:2006za,Sjostrand:2007gs} for showering. The dependence on astrophysical parameters is contained in the J-factor, defined as
\begin{equation}
J(\Delta \Omega) \equiv \int_{\Delta \Omega} \int_{l.o.s} \rho_{DM}^2(\mathbf{r}) \, dl\,d\Omega,
\end{equation}
which represents the integral of the squared dark matter density $\rho_{DM}(\mathbf{r})$ over the line of sight and over the solid angle $\Delta \Omega$ corresponding to a region of interest.  A common parameterization of the dark matter density in the Milky way is an Einasto profile~\cite{1965TrAlm...5...87E} 
\begin{equation}
\rho_{DM}(r) = \rho_0 \exp\left( -\frac{2}{\alpha}\left( \frac{r}{r_s} \right)^\alpha \right),\label{eq:Einasto}
\end{equation}
where $r$ is the distance from the galactic center.  A common set of values is $\alpha = 0.17$, $r_s=20\, \rm{kpc}$, and $\rho_0$ such that $\rho_{DM}(r_{\odot}) = 0.4\, \rm{GeV}\, \rm{cm}^{-3}$, where $r_{\odot} = 8.5\, \rm{kpc}$~\cite{Pieri:2009je,Silverwood:2014yza,Acharyya:2020sbj}.  We will specialize to this profile in our studies of CTA.  More cored profiles are more difficult to probe (see e.g.~\cite{Pierre:2014tra} for a related discussion for CTA).

The flux in Eq.~(\ref{eq:flux}) can be combined with the specifications of a given experiment to calculate the expected number of observed dark matter photons, $\mu_{DM}$:   
\begin{equation}
\mu_{DM} = T_{obs} \int_{E_1}^{E_2} \frac{d\Phi_{\gamma}}{dE} A_{eff}(E) \, dE.
\end{equation}
Here, $ T_{obs}$ is the duration of observation, $E_1$ and $E_2$ are the bounds of the energy range being observed, and $A_{eff}(E)$ is the effective area of an instrument as a function of the energy of the observed gamma ray.  This prediction can then be compared with data after accounting for astrophysical backgrounds.

For CTA, the dominant background across all energies comes from cosmic rays (CR).  
The expected CR background for CTA-South is available at~\cite{CTAArea}. 
For Fermi-LAT, the dominant background is the diffuse gamma ray emission from CR interactions with the interstellar medium and interstellar radiation field.  The diffuse gamma rays also constitute an important background for CTA.  Both experiments must also contend with localized sources, and one additional background for CTA will be the Fermi bubbles~\cite{Acharyya:2020sbj,Calore:2018sdx}. 

\section{Gamma Ray Spectra from Dark Matter Annihilation}

In this section, we present annihilation spectra for secluded dark matter. Many of our results, while derived in a supersymmetric framework, will also be applicable to non-supersymmetric setups.  We take dark matter to be a stable (scalar or fermion) particle in the secluded sector that annihilates %with thermal cross sections 
into lighter secluded sector particles.
%to achieve the correct relic density. 
The gamma ray spectrum is dominated by the continuum emission of photons from the visible sector decay products of the secluded sector annihilation products. Annihilations to a monochromatic $\gamma$-ray line, generally of interest for dark matter searches (e.g. see~\cite{Rinchiuso:2020skh} for an analysis of this case at CTA), are suppressed by the the portal coupling squared and are typically irrelevant.

For concreteness, we assume the fields in the secluded sector are charged under a $U(1)'$ symmetry, which is spontaneously broken. The portal between the secluded and visible sectors is the supersymmetric kinetic mixing portal, given by~\cite{Dienes:1996zr} 
\begin{equation}
\frac{\epsilon}{2} \int d^{2} \theta \,  W_{Y} W^{\prime} + h.c.  = \epsilon D_{Y} D^{\prime} - \frac{\epsilon}{2} F_Y^{\mu \nu}  F^{\prime}_{\mu \nu} + i \epsilon \tilde{B} \sigma^{\mu} \partial_{\mu} {\tilde{B}}^{\prime \, \dagger}
+ i \epsilon \tilde{B}^{\prime} \sigma^{\mu} \partial_{\mu} \tilde{B}^{\dagger},\label{eq:SUSYkinetic}
\end{equation}
where $W^{\prime}$ and $W_{Y}$ represent the chiral field strength superfields for the secluded sector $U(1)'$ and SM hypercharge, respectively. This leads to the typical kinetic mixing term between gauge field strengths, as well as a Higgs portal via the D-term coupling, and a gaugino mixing via the bino $\tilde{B}$ coupling with the secluded bino $\tilde{B'}$.\footnote{There is an additional soft mass term between the bino and secluded bino, $m_{\tilde{B} \tilde{B}^\prime }$ that can impact mixing in the gaugino sector.}

We will first consider the case where dark matter annihilates dominantly into secluded sector Higgs ($H'$) or gauge ($Z'$) bosons. This limit matches a minimal secluded sector containing only a $Z'$ boson, a Higgs boson $H'$, and a dark matter candidate, and our results overlap with previous studies of such minimal setups~\cite{Evans:2017kti,Hooper:2019xss,Profumo:2017obk}. However, this scenario also applies to supersymmetric cases where annihilations into secluded neutralinos are subdominant or kinematically forbidden. The couplings in Eq.~(\ref{eq:SUSYkinetic}) provide portals for both bosons to the SM, determining control the decays of secluded bosons into SM states. 

Next, we turn to annihilations into fermions in the secluded sector: the superpartners of the $H'$ and $Z'$, which mix to form Majorana neutralino states $\chi_{1}^\prime, \chi_{2}^\prime$.  If $\chi_{1}^{\prime}$ is the lightest fermion in the secluded sector, it decays to the MSSM through gaugino mixing (see e.g. Ref.~\cite{Barnes:2020vsc} for a recent realization). It is possible that the $\chi_{1}^{\prime}$ is itself the LSP, in which case it may only decay via a combination of neutralino mixing and RPV couplings.  As we will discuss, which RPV couplings are largest impacts the photon spectrum. Alternatively, if the LSP resides in the visible sector, gaugino mixing will induce $\chi_1'$ decays to the LSP and a SM boson (typically $Z$ or $h$). The LSP may either be stable (in which case it would itself provide a contribution to the dark matter density) or decay if R-parity is not conserved. While this discussion is framed in a supersymmetric context, the results are applicable to non-supersymmetric setups where dark matter annihilates into secluded sector fermions, which subsequently decay into SM fermions, possibly with intermediate cascade steps.   

When examining spectra, it is useful to bear in mind the relative importance of the soft and hard regions.   This depends on instrumental parameters. As we will see, for CTA, the sensitivity climbs rapidly with photon energy,  
hence CTA is more sensitive to harder spectra. Therefore, spectra with the largest peaks may not be best probed by CTA, especially if the peak is at low energy. Fermi, on the other hand, is sensitive to the majority of photons from weak scale annihilations, so the peak height is a better indicator of sensitivity.

\subsection{Annihilation to Secluded Sector Bosons}

First we consider cases where dark matter dominantly annihilates to a pair of secluded sector bosons, $Z^{\prime} H^{\prime}$, $Z^{\prime} Z^{\prime}$, or $H^{\prime} H^{\prime}$.  The relative importance of these channels is determined by both the spin of the dark matter and the origin of DM and gauge boson masses~\cite{Bell:2016uhg,Bell:2016fqf}. First, consider a Dirac fermion dark matter candidate, $\psi$. If $m_\psi$ and $m_{Z^{\prime}}$ are set by the vacuum expectation value (vev) of $H'$, the dominant annihilation channel will be $Z'H'$ if kinematically accessible. In other models where the mass generation mechanisms of the $\psi$ and $Z'$ are divorced from each other, as might be the case if DM possesses a vector-like mass or if the gauge boson receives its mass via the Stueckelberg mechanism, then a large gauge coupling or judicious charge assignments can produce annihilation dominantly to $Z^{\prime}Z^{\prime}$, see~\cite{Bell:2016uhg}. For fermionic dark matter, the $H'H'$ channel is $p$-wave suppressed and thus typically irrelevant for indirect detection; however, for scalar dark matter this is not the case, and this channel can be the most important. 

 The photon spectra from these annihilations is determined by the branching ratios of the secluded sector bosons to SM states. We assume a Higgs portal and a vector portal, 
 \begin{equation}
 \mathcal{L} \supset \xi |H'|^2 |H|^2 - \frac{\epsilon}{2} B^{\mu \nu } Z^{\prime}_{\mu \nu},
 \end{equation}
 where $H$ is the SM-like Higgs field, and $B_{\mu \nu}$ the SM hypercharge gauge field strength. If these portals are generated via the D-term of the supersymmetric kinetic mixing in Eq.~(\ref{eq:SUSYkinetic}), then in the Higgs decoupling limit $\xi\approx \frac{\eps}{2} g' g_Y \cos{2\beta}$, where $\tan{\beta}$ is the ratio of the MSSM Higgs vevs.   For our analysis, we assume only SM states are kinematically accessible, in which case this portal determines  the $H',\,Z'$ branching ratios.  For sufficiently heavy secluded sector bosons, it is conceivable that they might decay to heavier MSSM states, but we do not pursue this possibility further.
 
 \begin{figure}[t]
    \begin{minipage}{0.5\textwidth}
        \includegraphics[width=\textwidth]{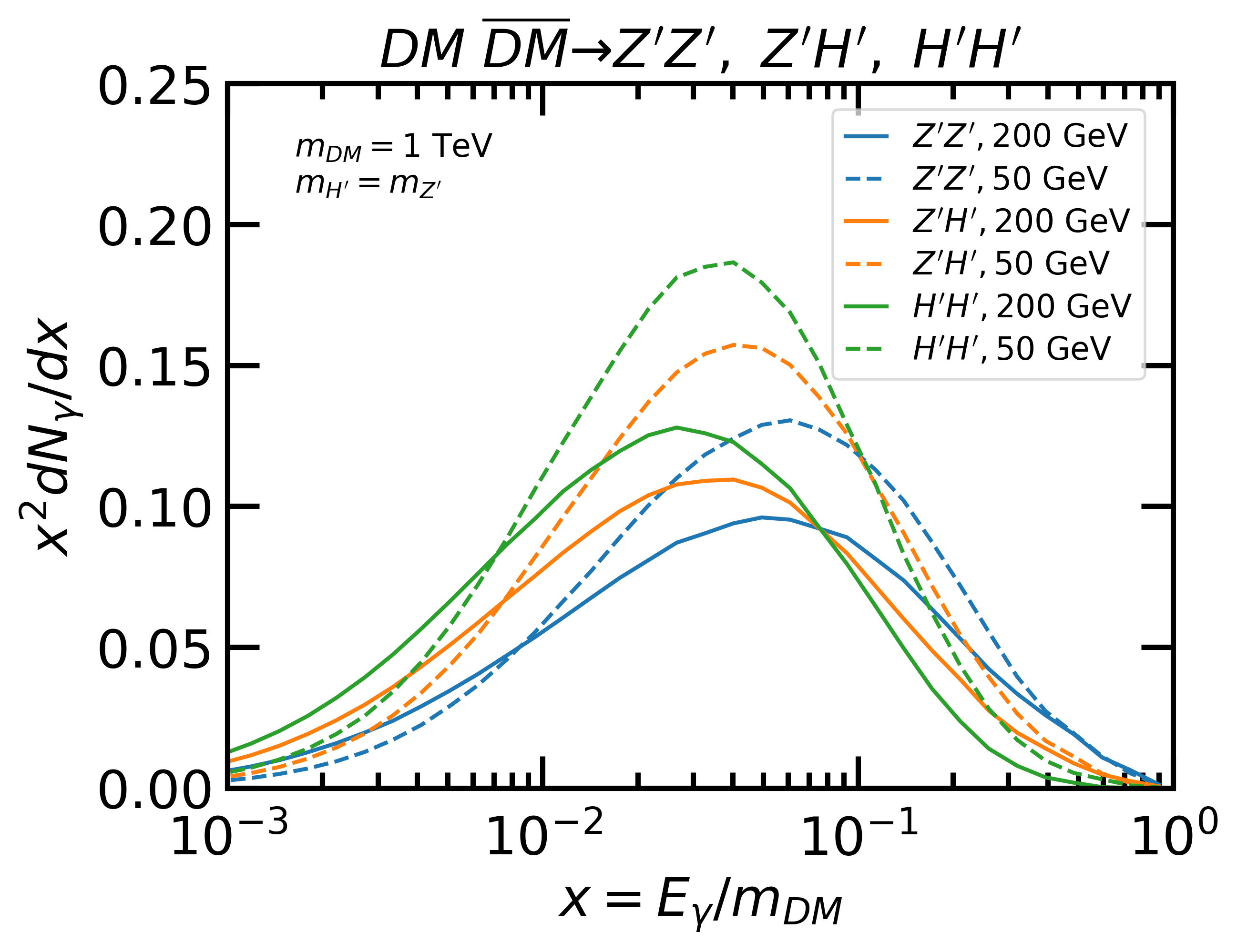}
    \end{minipage}%
    \begin{minipage}{0.5\textwidth}
        \includegraphics[width=\textwidth]{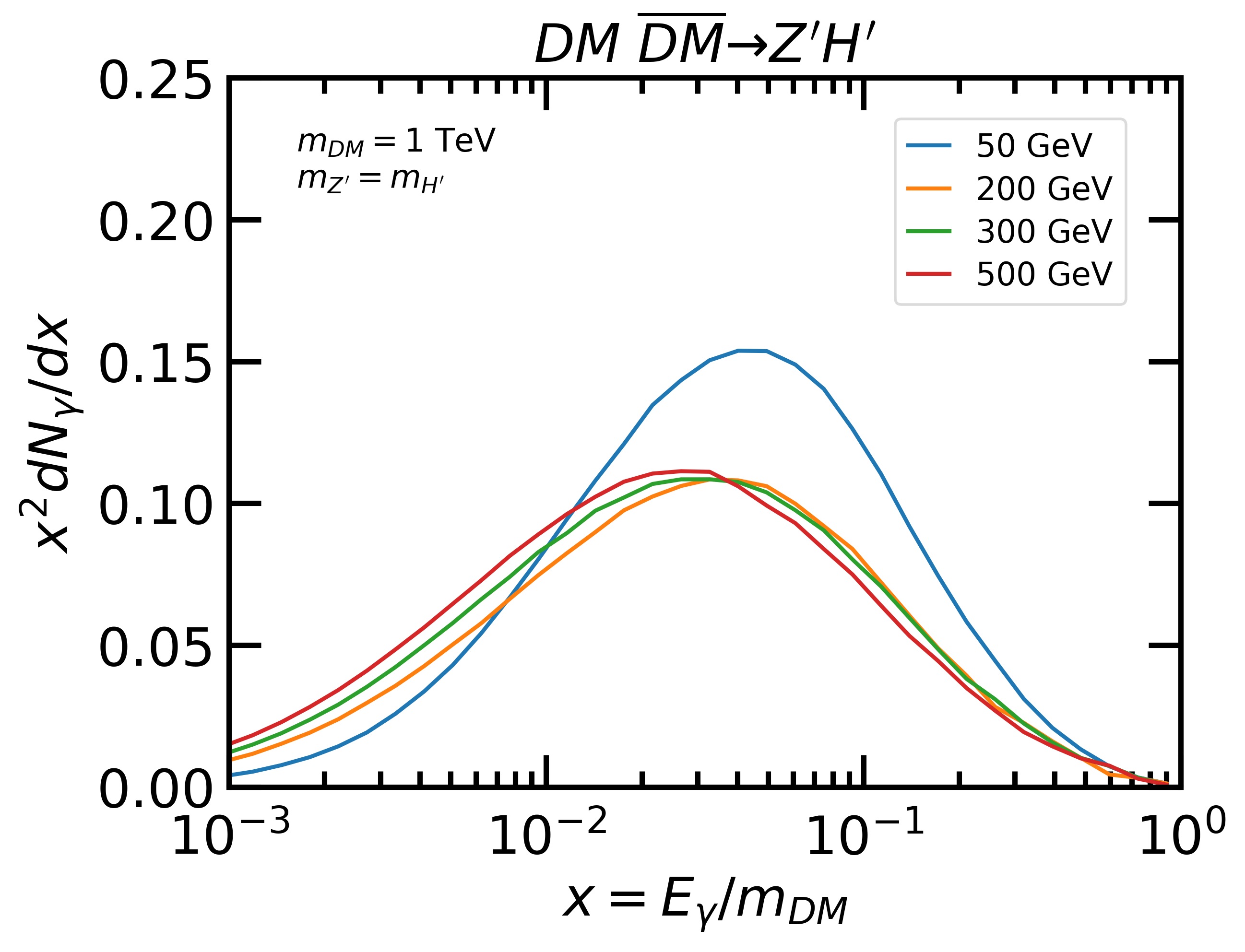}   
    \end{minipage}\ \\ 
    
    \begin{minipage}{0.5\textwidth}
        \includegraphics[width=\textwidth]{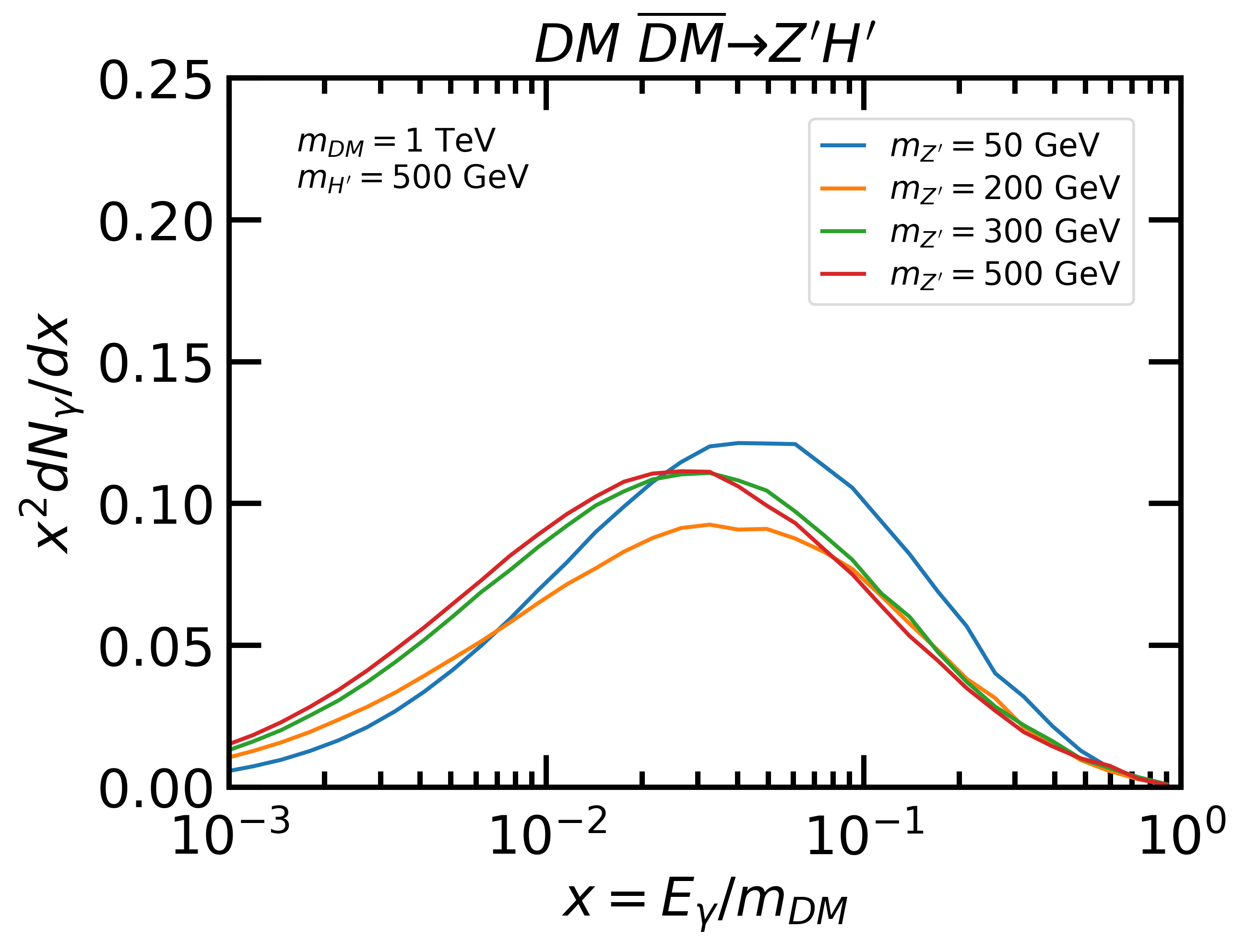}
    \end{minipage}%
    \begin{minipage}{0.5\textwidth}
        \includegraphics[width=\textwidth]{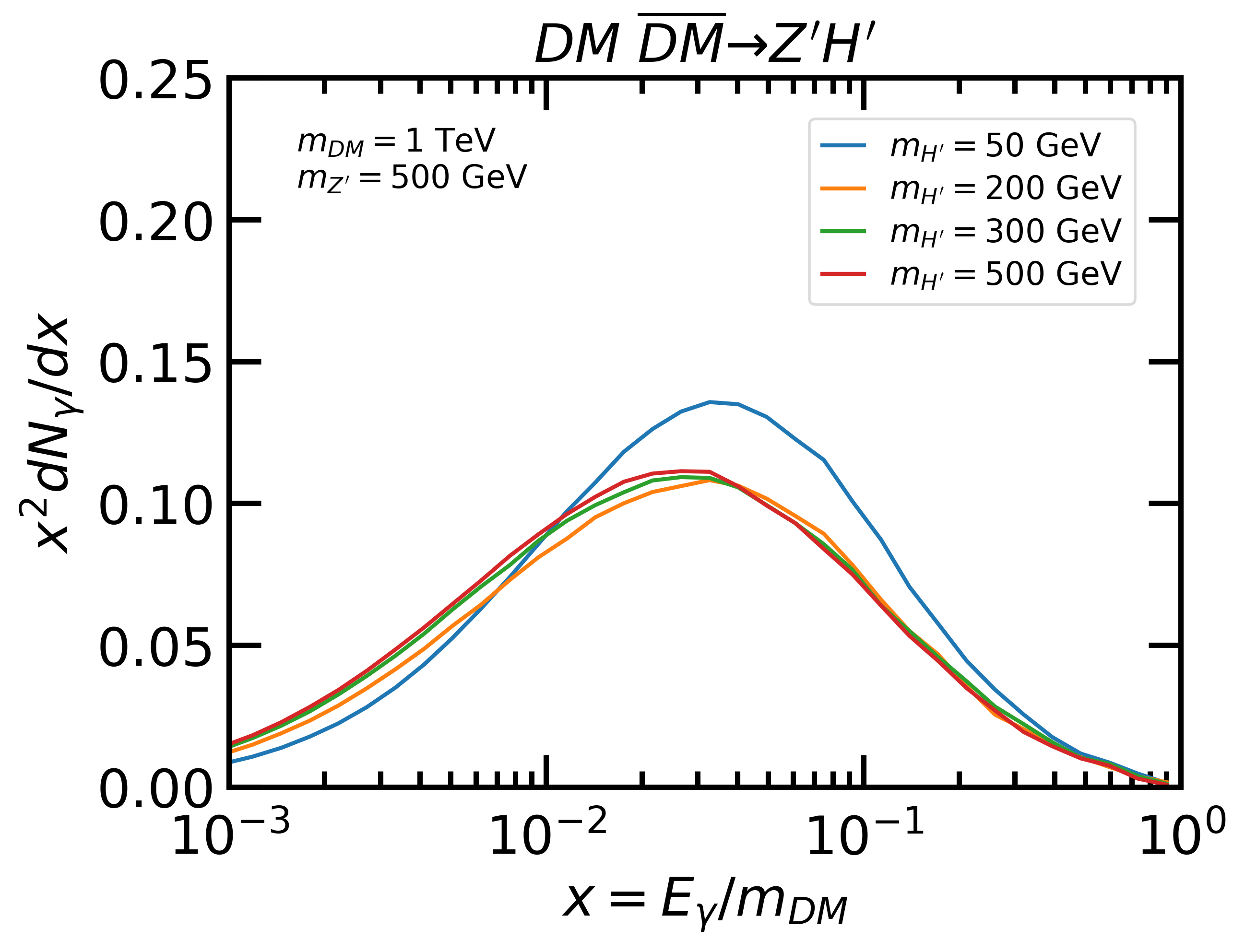}   
    \end{minipage}
    \caption{\textbf{Top Left:} Photon spectra for (scalar or fermion) DM annihilation to combinations of $Z',\,H'$ final states in the degenerate case $m_{Z'}=m_{H'}$. 
    \textbf{Top Right:} Photon spectra of the final state $Z'H'$ for several $m_{Z'}=m_{H'}$. \textbf{Bottom Left:} Photon spectra for $Z'H'$ in the non-degenerate case. Fixed $m_{H'}$ and varying $m_{Z'}$.
    \textbf{Bottom Right:} Fixed $m_{Z'}$ and varying $m_{H'}$. }
    \label{fig:ZH_ZZ}
\end{figure}
 
 We show photon spectra from dark matter annihilation to various combinations of $Z'$ and $H'$ in FIG.~\ref{fig:ZH_ZZ}.  For low masses (e.g. $m_{H'}=50$ GeV, as shown in the figure), the $H'$  dominantly decays  to $b\bar{b}$. If sufficiently heavy, the dominant decay channel is $WW$, and the multiplicity of the hardest photons from $H'$ decays is significantly lower (compare the green solid and dashed curves in the top left panel of FIG.~\ref{fig:ZH_ZZ}), resulting in weaker indirect detection bounds.
  For the $Z'$ boson, if $m_{Z'}\ll m_{Z}$, the $Z'$ couples predominantly to electric charge $Q$, while for $m_{Z'}\gg m_{Z}$, it mainly couples to hypercharge $Y$. In either case, decays to up-type quarks and charged leptons have the largest branching ratios.  The up-type quarks provide the largest contribution to the photon spectrum, and their spectra are harder than either of the dominant final states in $H'$ decay.  In FIG.~\ref{fig:ZH_ZZ}, this can be observed on the top left panel, where the $Z^{\prime} Z^{\prime}$ spectra have more support at large $x$ than their $Z^{\prime} H^{\prime}$  and $H^{\prime} H^{\prime}$ counterparts. Variations of the DM mass only mildly affect $dN/dx$, especially if it is much heavier than its annihilation products~\cite{Elor:2015bho}. For cases in which the boson decays to quarks, the boson's mass can have a significant effect on the spectrum, where lighter masses generate fewer but more boosted pions.  This results in harder spectra.    

In general, $m_{Z^\prime}\neq m_{H^\prime}$ (in supersymmetric models, these masses are identical up to loop corrections). When kinematically allowed, $H'$ will decay to pairs of $Z'$ rather than directly to SM states. However, for $m_{H'} > 2 m_W$, this does not result in an increased number of cascade steps. We examine the potential impact of this new decay mode in the bottom left panel of FIG.~\ref{fig:ZH_ZZ}. Everywhere in this panel $H'$ decays dominantly to either $Z'Z'$ (if kinematically open) or $WW$. Thus differences in spectra arise from the difference in mass and decay patterns between the $Z'$ and $W$. For instance, the $W$ produces roughly $50\%$ more quarks and half the taus as a $200$ GeV $Z'$ ~\cite{Escudero:2017yia}, resulting in a similar multiplicity of hard photons but fewer soft photons (compare green and orange curves). For decays to quarks, a lighter $Z'$ results in relatively hard spectra.  This effect, along with the increased branching ratio of $Z'$ below $m_Z$ to quarks at the expense of neutrinos, explains the relative hardness of the $m_{Z'}=50$ GeV curve relative to the others. In the bottom right panel, as  $m_{H'}$ varies, no new $Z^{\prime}$ decay channels open up, so the change in the $H^{\prime}$ branching ratios due to the $WW$ threshold is the dominant effect. This effect is not significant for the masses shown due to the $Z'$ and $H'$ having relatively similar spectra, but we will see cases where this kinematic effect is important (see FIG.~\ref{fig:RPV Mass vary}).

\subsection{Annihilation to Secluded Sector Neutralinos}\label{sec:secludedNeut}
We now turn to the case where DM primarily annihilates to the secluded neutralinos $\chi^{\prime}$.  We concentrate on annihilations to the lighter neutralino $\chi_{1}^{\prime}$.  In principle, annihilations to $\chi_2^{\prime}$ are also possible, which could lead to hidden sector cascades and softer spectra, typically more difficult to probe. We divide our discussion according to whether the secluded $\chi^{\prime}_1$ or the visible sector neutralino $\chi_1$ is the LSP.

\subsubsection{LSP in the visible sector}

\begin{figure}[t]\hspace{0.2in}
    \centering
    \begin{minipage}{0.4\textwidth}
		\includegraphics[width=1.0\textwidth]{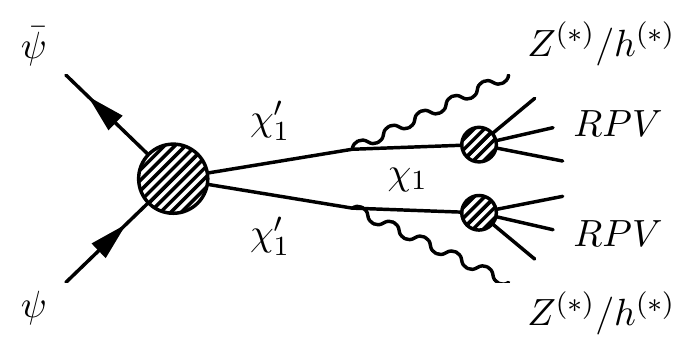}
    %\label{fig:RPV Diagram}
    \end{minipage}%
    \begin{minipage}{0.6\textwidth}
        \includegraphics[width=0.8\textwidth]{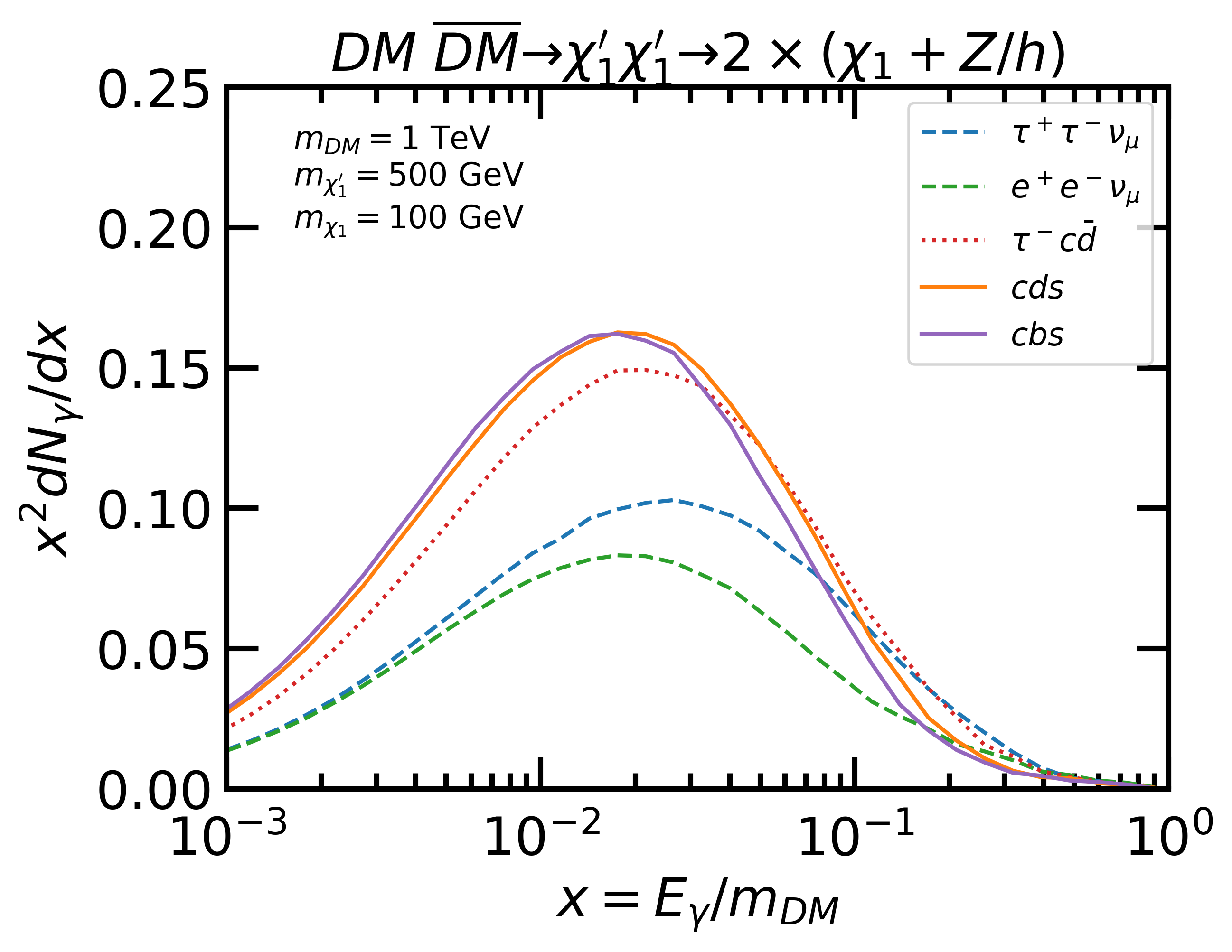}
%        \subcaption*{}
        %\label{fig:RPV Plot}
    \end{minipage}
    \caption{
    \textbf{Left:} DM annihilation through a neutralino cascade. ``RPV" indicates the  three fermion final state from $\chi_1$ RPV decay, which differs based on the dominant  RPV coupling. If no RPV interactions are present, the $\chi_1$ is stable and carries away energy. Here we assume sfermions are sufficiently heavy that even if off-shell,  $Z^{\ast}$, or $h^{\ast}$ dominate the decay.
    \textbf{Right:} The resulting spectra for specific choices of nonzero RPV couplings (all others set to zero).
    Here we take BR$(\chi'_1 \to \chi_1 h)=95 \%$, and the rest to $\chi_1 Z$. }
    \label{fig:RPV}
\end{figure}

Suppose that the LSP is an MSSM neutralino $\chi_1$.  If kinematically accessible, $\chi_1^\prime$ will decay to it and an accompanying SM boson\footnote{We do not consider decays to accompanying secluded sector bosons.  In the model considered in~\cite{Barnes:2020vsc}, the lightest secluded neutralino is always lighter than these states.} ($Z$ or $h$) via kinetic mixing between the bino and secluded bino. The LSP is either stable, in which case it constitutes a separate portion of the DM abundance, or it may decay via one of the RPV interactions
\begin{equation}
    W_{RPV} = \frac{1}{2}\lambda_{ijk} L_i L_j \bar{E}_k+\lambda^\prime_{ijk} L_i Q_j \bar{D}_k+\frac{1}{2}\lambda^{\prime \prime}_{ijk} \bar{U}_i \bar{D}_j \bar{D}_k\,.
\end{equation}
These couplings induce 3-body decays of the LSP via off-shell squarks and sleptons (see FIG.~\ref{fig:RPV}, left panel). 

In FIG.~\ref{fig:RPV} (right panel), we show characteristic spectra from this annihilation process for different choices of the dominant RPV coupling.  The three dash types denote the three sets of trilinear RPV couplings, while variations in the colors correspond to different generations.  The differences in the spectra reflect the underlying photon spectra each SM final state generates. The tau leptons give a harder spectra than the quarks but produce fewer low energy photons~\cite{Cirelli:2010xx}.  Electrons (and muons) produce far fewer photons overall.   Differences between the final states corresponding to  different RPV couplings are smaller than might otherwise be expected because of the non-negligible contribution from the decays of the final state $Z$/$h$. The relative branching ratio into the SM bosons depends on the detailed spectrum of the MSSM, the hidden sector, and $\epsilon$. If $\chi_1$ is dominantly bino, then $\chi'_1$ predominantly decays to $\chi_1 h$. In each figure we specify the branching ratios used to obtain the spectra.

\begin{figure}[t]
%    \centering
    \begin{minipage}{0.5\textwidth}
        \includegraphics[width=\textwidth]{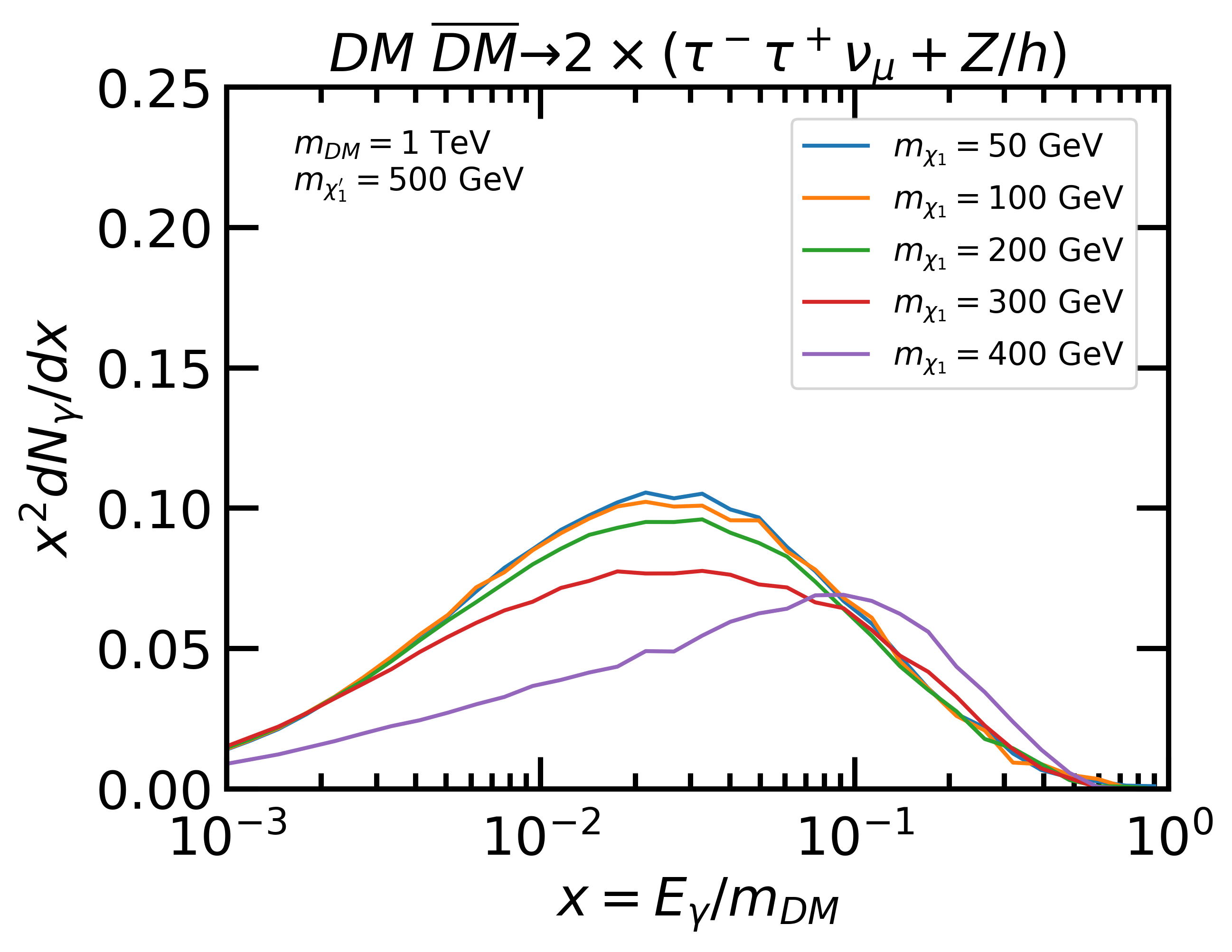}
    \end{minipage}%
    \begin{minipage}{0.5\textwidth}
    \includegraphics[width=\textwidth]{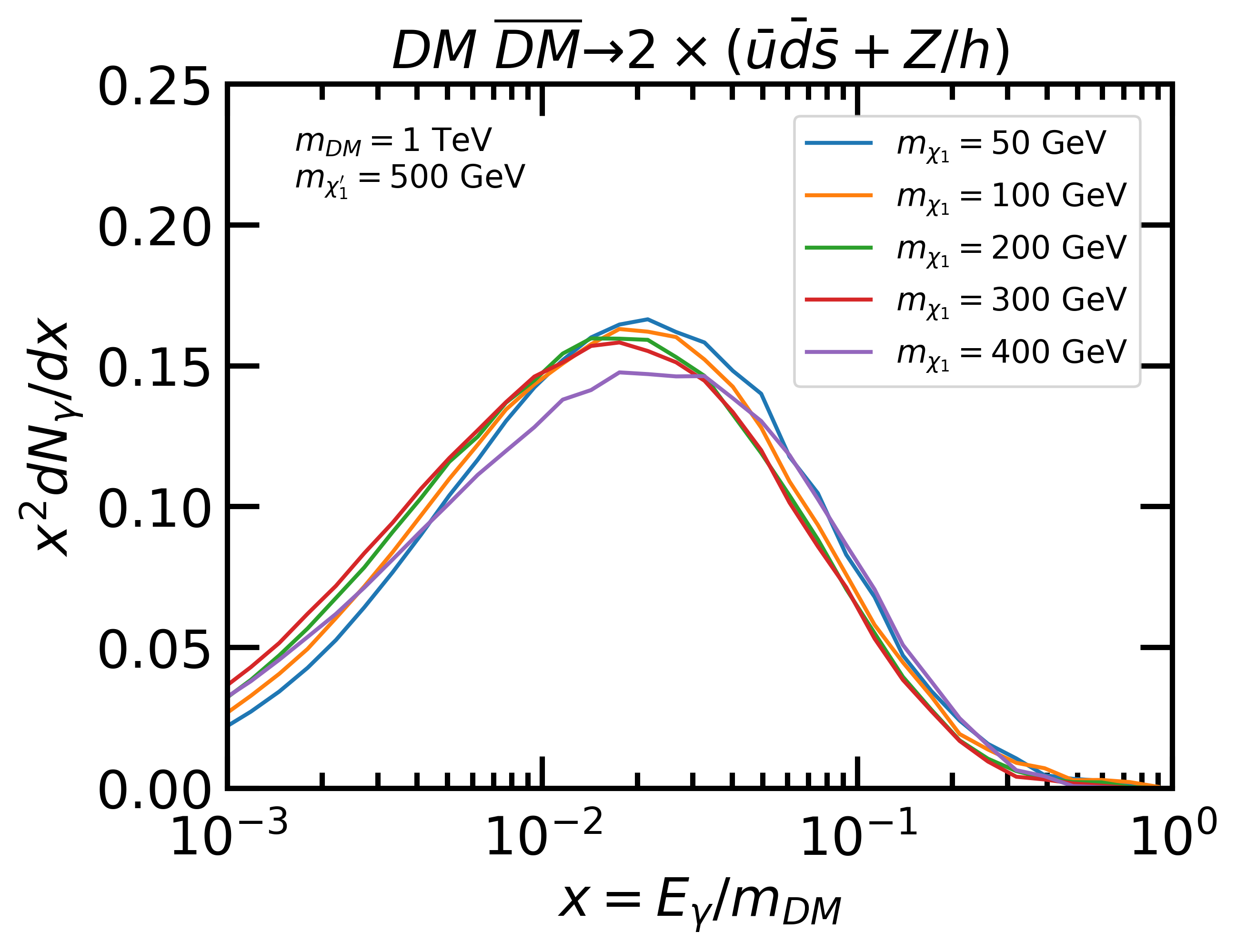}
    \end{minipage}
    \caption{
    The photon spectra for various LSP masses. For all curves BR$(\chi'_1 \to \chi_1 h) \gtrsim 99 \%$, except the $400$ GeV curve, for which kinematics forces $100\%$ BR to $\chi_1 Z$. \textbf{Left:} We set all RPV couplings to be zero, except $\lambda_{323}=-\lambda_{233}\neq0$.
    \textbf{Right:} We instead make only $\lambda^{\prime\prime}_{112}=-\lambda^{\prime\prime}_{121}\neq0$. }
    \label{fig:RPV Mass vary}
\end{figure}

Varying the mass splitting between the visible and secluded sector neutralinos can have a pronounced effect on the spectrum, see FIG.~\ref{fig:RPV Mass vary}. As the splitting decreases, the highest energy photons come exclusively from the RPV decay. 
 This effect is most clear in the left panel, due to the hard photons from $\tau$ decay. In the right panel, with pure quark RPV coupling, the photon spectra are less sensitive to changes in mass splittings due to the similarity between the photon spectra of $Z$ bosons and quarks.

\subsubsection{LSP in the secluded sector}

\begin{figure}[t!]
    \begin{minipage}{0.5\textwidth}
        \includegraphics[width=1\textwidth]{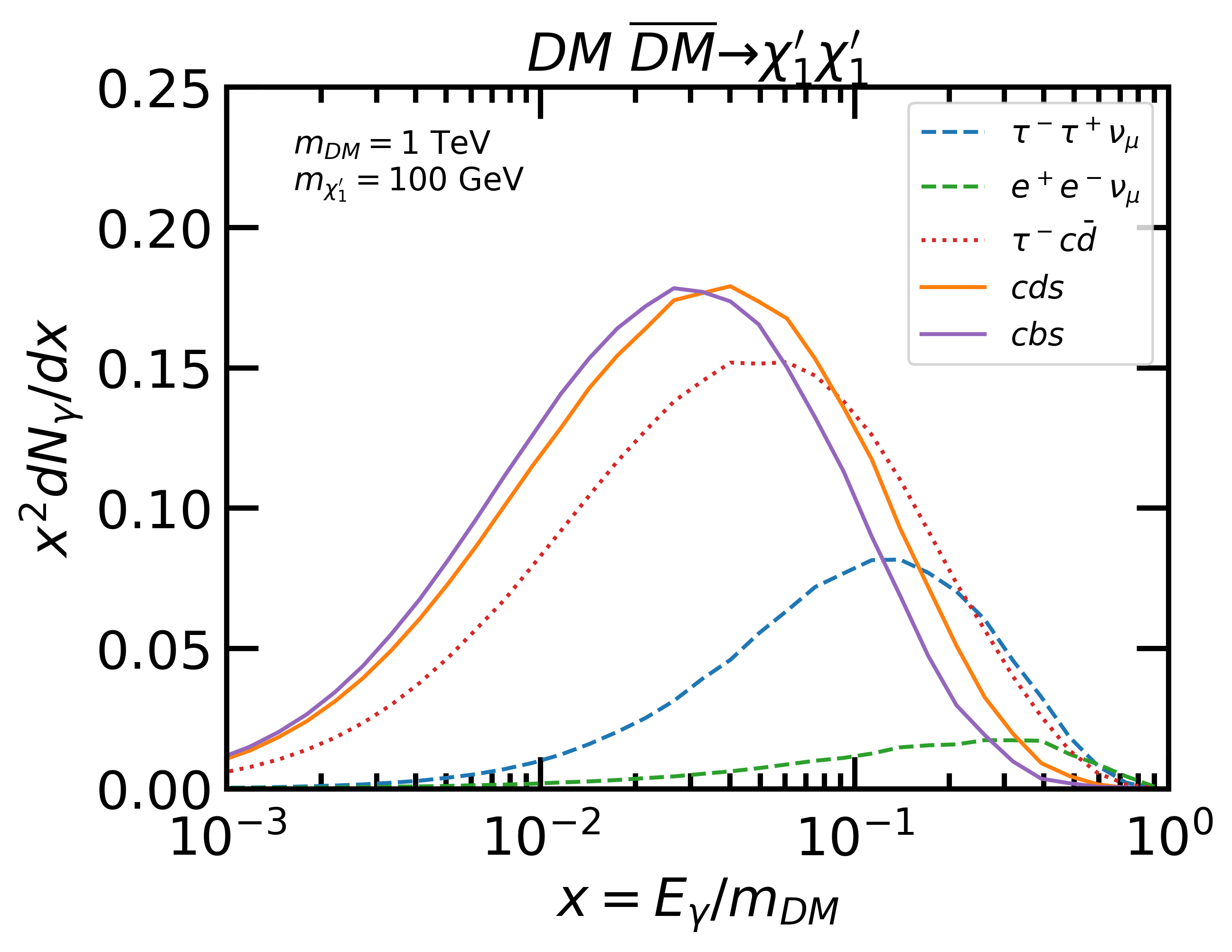}
    \end{minipage}%
    \begin{minipage}{0.5\textwidth}
        \includegraphics[width=1\textwidth]{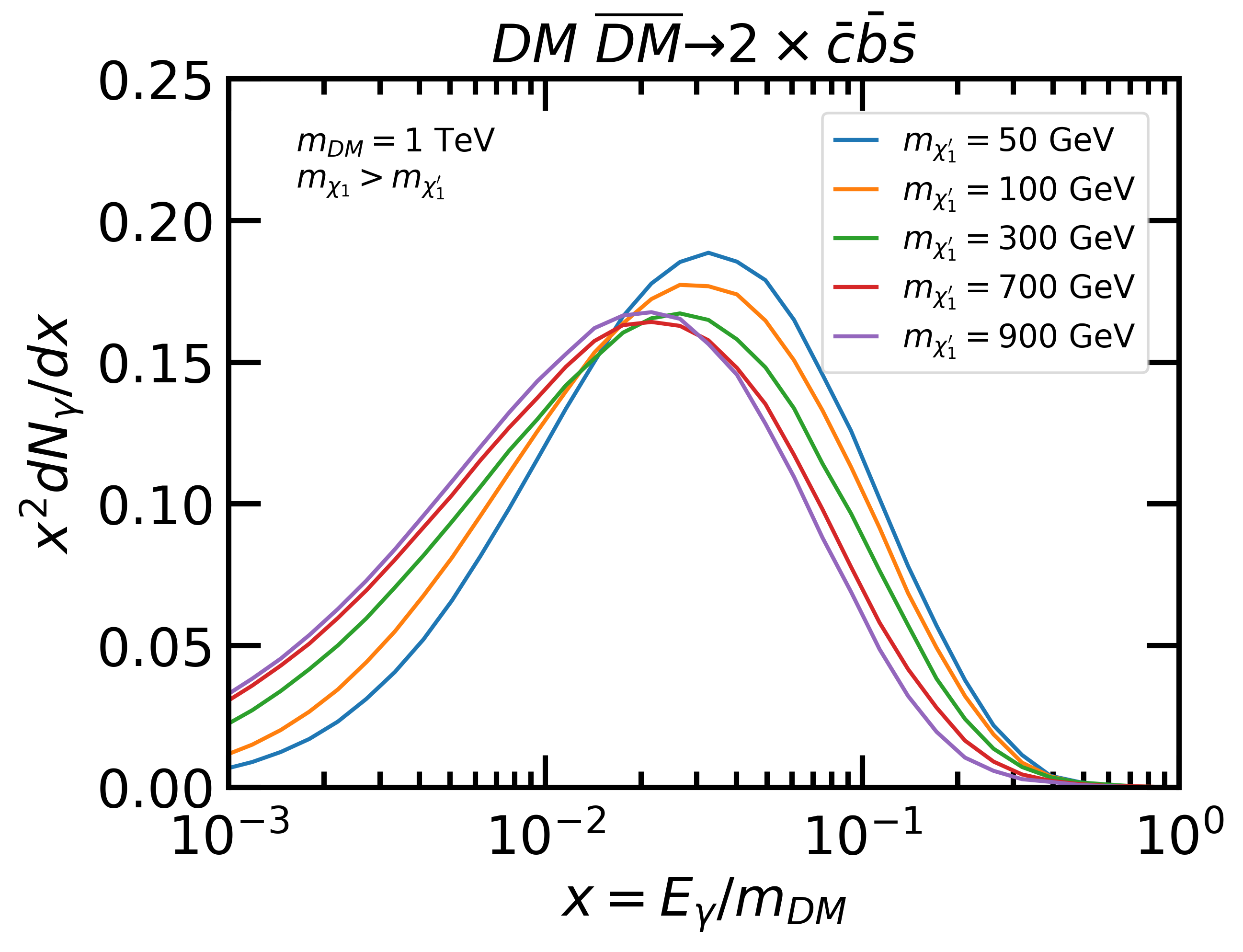}
    \end{minipage}
    \caption{\textbf{Left:} The photon spectra for DM annihilation to $\chi_1^\prime$, here the LSP, shown for multiple potential RPV mediated $\chi_1^\prime$ decays. \textbf{Right:} We scan over the mass of the $\chi_1'$, fixing the RPV coupling to $\lambda^{\prime \prime}_{223}$.}
    \label{fig:direct_chi-1}
\end{figure}

If the $\chi_1^{\prime}$ is lighter than its MSSM counterparts, it decays directly to the SM via RPV couplings. This decay rate is suppressed by $\epsilon^2$ relative to the MSSM neutralino decay rate. The corresponding lifetime of $\chi_1'$ is 
\begin{equation}
    \tau \approx 1 \textrm{ s} \times \left(\frac{10^{-4}}{\epsilon}\right)^2 \left(\frac{0.01}{\lambda}\right)^2 \left(\frac{50 \; {\textrm{GeV}}}{m_{\chi_1^\prime}}\right)^5 \left(\frac{\tilde{m}}{10\ {\textrm{TeV}}}\right)^4, \label{eqn:BBN}
\end{equation}
where $\tilde{m}$ represents the sfermion mass scale. The freeze-out abundance of the  $\chi_1^\prime$ is significant, and we have checked that co-annihilations with the SM bath (which are suppressed by $\epsilon$) are not sufficient to reduce the abundance to avoid bounds from decays after Big Bang Nucleosynthesis (BBN)~\cite{Poulin:2016anj,Dienes:2018yoq}. Hence this lifetime is strictly bounded by BBN. Nevertheless, Eq.~(\ref{eqn:BBN}) yields a large parameter space in which $\chi_1^\prime$ RPV decay does not conflict with BBN. In FIG.~\ref{fig:direct_chi-1} (left panel) we show the analogue of FIG.~\ref{fig:RPV} (right panel), where now the $\chi_1'$ instead decays directly via RPV. Since there is no additional SM boson production,  the differences between leptonic and baryonic RPV decays are more pronounced. For all SM annihilation products except $e$'s and $\mu$'s, the dominant contribution to the photon spectrum is from hadronic showering. In this panel we fixed the two relevant mass scales. 
The invariant mass of the parent particle whose decay initiates hadronic showering may have a significant effect, particularly as one approaches the QCD scale. This is shown in FIG.~\ref{fig:direct_chi-1} (right panel).  Lighter masses typically correspond to harder spectra, resulting in more stringent projected bounds on RPV decays into quarks from CTA.

\begin{figure}[t!]
    \centering
    \includegraphics[width=0.625\textwidth]{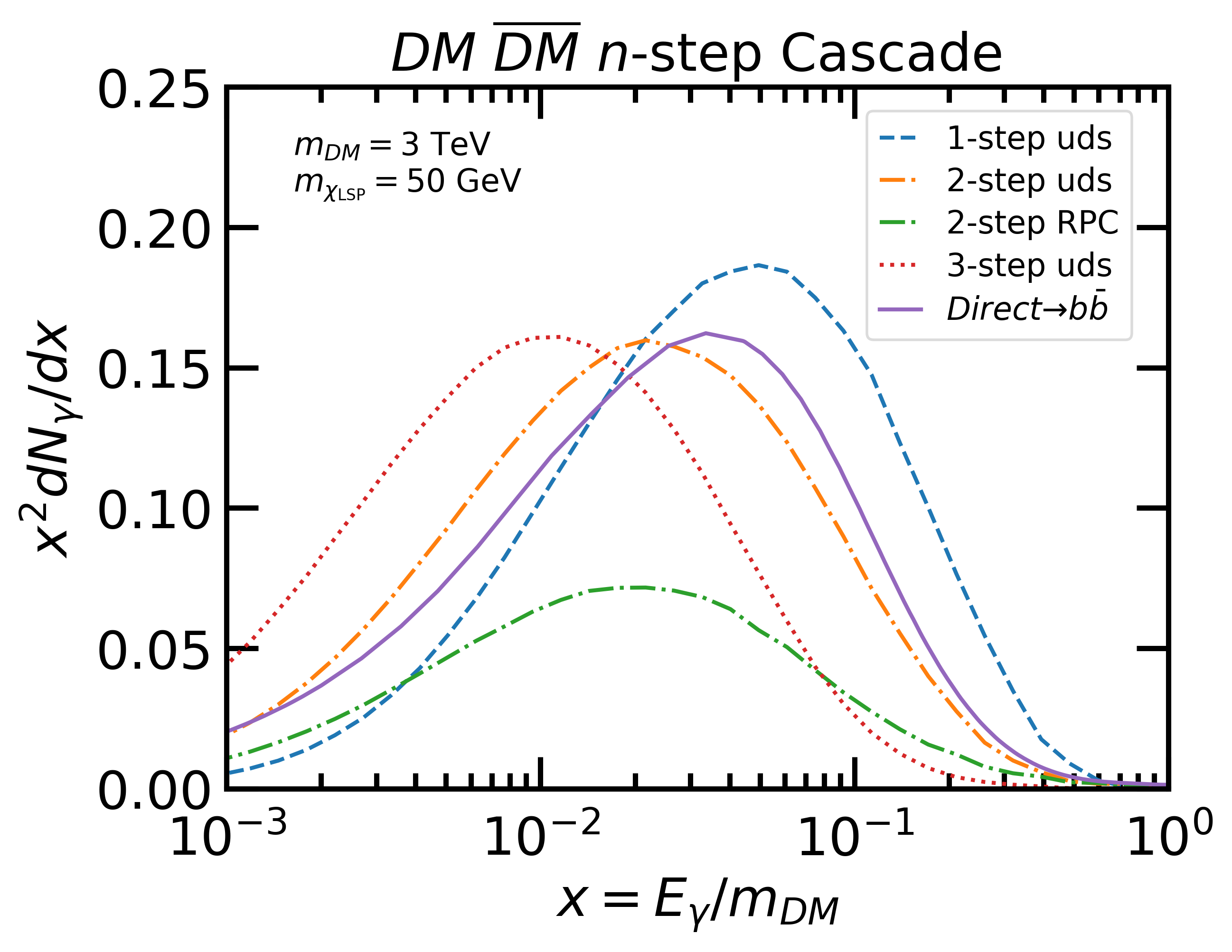}
    \caption{Photon spectra for a variety of cascades. 
    %1-step denotes DM annihilation to LSP $\chi'_1$, which decays via RPV. Similarly, 2-step denotes annihilation to $\chi'_1$, but with BR$(\chi'_1 \to \chi_1 h)=64 \%$ and the rest to $\chi_1 Z$. The  LSP, $\chi_1$, can be either stable (RPC) or decay $(uds)$. For 3-step processes, DM first annihilates to $Z'H'$, each boson decays to a pair of $\chi'_1$, which then follows the 2-step cascade. 
    {\bf 1-step:} $DM\overline{DM}\to\chi_1'\chi_1'$; $\chi_1'\to$ RPV. {\bf 2-step:} $DM\overline{DM}\to\chi_1'\chi_1'$; BR$(\chi'_1 \to \chi_1 h)=64 \%$, rest to $\chi_1 Z$; LSP $\chi_1$ stable (RPC) or decays ($uds$).
    {\bf 3-step:} $DM \overline{DM} \to Z' H'$; $Z'/H' \to \chi'_1\chi'_1 $; $\chi'_1$ decays as in the 2-step cascade.
    For the RPV cases, we assume $\lambda''_{112}$ $(uds)$ dominates.  For all cases we fix the LSP mass to $50$ GeV. The mass of the $\chi'_1$ when it is not the LSP are $1500$ GeV (2-step), or $1125$ GeV (3-step); the 3-step $\chi'_1$ differs to allow $Z' \rightarrow \chi_1^{\prime} \chi_1^{\prime}$. 
    %with $m_{Z'}<m_{DM}$ must decay to a pair of $\chi'_1$.  
    For comparison, a spectrum with direct annihilations to $b\bar{b}$ is also shown.}
    \label{fig:Photon Step Comparison}
\end{figure}

We compare multi-step cascades for the RPV and $R$-parity conserving (RPC) cases in FIG.~\ref{fig:Photon Step Comparison}.  While secluded sector annihilation products may decay directly to the SM or an MSSM state, a secluded sector with sufficiently large mass splittings allows for cascades with an extra step, such as $\bar{\psi}\psi \to Z'H'$, with subsequent decays such as $Z' \to \chi_{1}^\prime\chi_{1}^\prime$.   These spectra are in agreement with previous studies of multi-step cascade annihilations of secluded sector dark matter~\cite{Elor:2015tva}. If the MSSM does not contain RPV couplings, then $\chi_1$ is stable and represents an additional DM component. In this case, the reduced relic abundance of the secluded DM will suppress indirect detection signals. Furthermore, the photon spectrum from each secluded sector annihilation will also be suppressed; for example, if $\chi_{1}^{\prime} \rightarrow \chi_{1} (Z/h)$, only the decays of the $Z/h$ contribute to the photon spectrum. This is shown by the green dashed curve in FIG.~\ref{fig:Photon Step Comparison}, and is clearly subdominant to its RPV counterpart in orange. For details on the cascade chains, see caption.

\section{Fermi Bounds and CTA projections}

Using the photon spectra from the previous section, we now derive bounds on dark matter annihilation cross sections from Fermi-LAT and projected sensitivity from CTA. But before doing so, we first discuss present-day cross sections consistent with the observed thermal relic abundance. This will make the significance of the derived bounds more apparent.

\subsection{Benchmark Cross Sections}\label{sec:sommerfeld}

The early universe annihilation cross section that achieves the relic density for weak scale dark matter (non self-conjugate) is $\langle \sigma v \rangle_{thermal}  \simeq  4.3  \times 10^{-26}$ cm$^{3}$ sec$^{-1}$. For $s$-wave annihilation \textemdash ~the case with a chance of being observed at indirect detection experiments \textemdash ~the naive expectation is a present-day cross section also equal to this value.

However, an important consideration in these models is the Sommerfeld effect.  In the presence of light mediators, the present-day annihilation cross section can be enhanced relative to its early Universe counterpart due to the low velocity of dark matter today. Thus, even with a thermal history, the cross section for indirect detection signals can be in excess of $\langle \sigma v \rangle_{thermal}$.  However, the maximum size of this mismatch is limited, as emphasized in Ref.~\cite{Feng:2010zp}: For a too-large at present day Sommerfeld effect, a nontrivial Sommerfeld effect will also be present at the time of freeze-out, reducing the relic density to an unacceptably low value. 

Using the {\tt DRAKE} code~\cite{Binder:2021bmg}, we calculate the relic abundance using the simple Boltzmann equation that assumes kinetic equilibrium between dark matter and the thermal bath.\footnote{It is known~\cite{Feng:2010zp,Binder:2017rgn,vandenAarssen:2012ag} that early kinetic decoupling can occur in cases where the Sommerfeld effect is appreciable, especially near resonance; however, the details of kinetic decoupling are model dependent, and can depend, e.g., on the value of the portal couplings that connect the two sectors.} 
We use present day s-wave cross sections as determined by the formula
\begin{equation}
    \langle \sigma v \rangle = S \langle \sigma v \rangle_{0},~~~
    S=\frac{\pi}{\epsilon_v} 
    \frac{\sinh\left(\frac{12  \epsilon_v}{\pi \epsilon_{\phi}}\right)}
    {\cosh{\left(\frac{12  \epsilon_{v}}{\pi \epsilon_{\phi}} \right)} -
    \cos{\left(
    2 \pi
    \sqrt{\frac{6}{\pi^2 \epsilon_{\phi}} - \left(\frac{6 \epsilon_{v}}{\pi^2 \epsilon_{\phi}}\right)^2} \right)}},
    \label{eq:Sommerfeld}
\end{equation}
where $S$ is the Sommerfeld enhancement factor as determined via the Hulth\'en potential approximation~\cite{Cassel:2009wt,Slatyer:2009vg}, with $\epsilon_{\phi} \equiv m_{\phi}/(\alpha_{\phi} m_{DM})$ for a massive mediator with mass $m_{\phi}$ that couples to dark matter with an effective coupling $\alpha_{\phi}$ that appears in the non-relativistic potential, and $\langle \sigma v \rangle_{0}$ is the tree level $s$-wave cross section. 
Here $\epsilon_v$ is defined in terms of the relative velocity $v$ as  $\epsilon_{v}\equiv v/(2\alpha_{\phi})$. Scalar Sommerfeld enhancement has been discussed, e.g., in~\cite{MarchRussell:2008yu,Hryczuk:2011tq}.

For the $Z'H'$ annihilation channel, we use the relevant cross section  $ \langle \sigma v \rangle_{0}$  from~\cite{Bell:2016uhg, Barnes:2020vsc}.  
\begin{equation}
\langle \sigma v \rangle_{Z^{\prime} H^{\prime}}= \frac{ \lambda^4}{64 \pi m_{DM}^2} \frac{(1-\eta_{Z^{\prime}})^{1/2} (64 - 128 \eta_{Z^{\prime}}+ 104 \eta_{Z^{\prime}}^2 - 30 \eta_{Z^{\prime}}^3 + \eta_{Z^{\prime}}^4 +\eta_{Z^{\prime}}^5)} {(2-\eta_{Z^{\prime}})^2 (4-\eta_{Z^{\prime}})^2},\label{eq:ZH}
\end{equation}
with $\eta_{Z^{\prime}}=m_{Z^{\prime}}^2/m_{DM}^2$. We expect the Sommerfeld enhancement from $H'$ exchange to dominate, 
hence $\alpha_{\phi}$ is a function of the relevant Yukawa coupling ($\lambda/\sqrt{2}$), $\alpha_{\phi} = \lambda^2/( 8 \pi)$.

For $Z^{\prime} Z^{\prime}$, we have
\begin{equation}
 \langle \sigma v \rangle_{Z^{\prime} Z ^{\prime}} = \frac{g^4}{ 16 \pi m_{DM}^2}  \frac{(1- \eta_{Z^{\prime}})^{3/2}}{(1-\eta_Z{^{\prime}}/2)^2}.\label{eq:ZZ}
 % LEANE and Bell eqn. 6.2
\end{equation}
 This cross section corresponds to a vector-like fermion, see e.g.~\cite{Bell:2016uhg}. In this case, $m_\phi=m_{Z'}$ and $\alpha_{\phi}=g^2/(4\pi)$ for Eq.~(\ref{eq:Sommerfeld}).

Finally, for the $H^{\prime} H^{\prime}$ final state, we consider the toy model
\begin{equation}
  V=\tilde{m}_{S}^2 |S|^2 + \lambda |S|^2 |H^{\prime}|^2 + \frac{\lambda^{\prime}}{2} \left( |H^{\prime}|^2 - \frac{v^{\prime 2}}{2}\right)^2,
    \end{equation}
where $S$ represents scalar dark matter. In the special case where the $S$ mass arises entirely out of $\langle H^{\prime} \rangle$, i.e. $\tilde{m}_{S}=0$,  the $s$-wave annihilation cross section is given by 
\begin{equation}
   \langle \sigma v \rangle_{H^{\prime} H^{\prime}} = 
   \frac{\lambda^2}{16 \pi  m_{S}^2} \sqrt{1-\eta_{H^\prime}} \frac{\left( 4 -2 \eta_{H^\prime} + \eta_{H^\prime}^2 \right)^2}{ (4-\eta_{H^\prime})^2 (2-\eta_{H^\prime})^2},   \label{eq:H'H'AnnSpecial}
\end{equation}
with $m_{S}=\sqrt{\lambda} v^{\prime}/\sqrt{2}$, $\eta_{H^{\prime}}\equiv m_{H^{\prime}}^2/m_{S}^2$, and  $m_{H^{\prime}}=\lambda' v'$.
The general case for $\tilde{m}_{S}\neq 0$  is:
\begin{equation}
       \label{eq:H'H'AnnGeneral}
       \langle \sigma v \rangle_{H^{\prime} H^{\prime}} = \frac{\lambda^2}{16 \pi m_{S}^2} \sqrt{1-\eta_{H^\prime}} \frac{\left( 4 -4 \eta_{H^\prime} \lambda/\lambda^\prime + \eta_{H^\prime}^2(\lambda/\lambda^\prime-1) \right)^2}{ (4-\eta_{H^\prime})^2 (2-\eta_{H^\prime})^2}.
\end{equation}
The previous expression is recovered for $\eta_{H'}=2 \lambda'/\lambda$.
In Eq.~(\ref{eq:Sommerfeld}), $m_\phi=m_{H'}$, and $\alpha_{\phi} = \lambda/(8\pi)$ for Eq.~(\ref{eq:H'H'AnnSpecial}) or $\alpha_{\phi} = \lambda^2\eta_{H'}/(16 \pi\lambda^{\prime})$ for the general case in Eq.~(\ref{eq:H'H'AnnGeneral}).  In the plots that follow in the next section, we specialize to the case of Eq.~(\ref{eq:H'H'AnnSpecial}).

Despite variations in the detailed forms of the cross sections listed above, we note that the thermal relic cross sections are quite close to each other.  When the Sommerfeld effect is unimportant, this cross section is simply $\langle \sigma v \rangle_{thermal}$.  For regimes where the Sommerfeld effect is important, i.e. in the limit of a light mediator, $\eta\to 0$, the cross sections for annihilation to $Z'Z'$, $Z'H'$, and $H'H'$ (Eqs.~(\ref{eq:ZH}),(\ref{eq:ZZ}), and (\ref{eq:H'H'AnnSpecial})) converge to the form $\langle \sigma v \rangle = \frac{\pi \alpha_{\phi}^2}{m_{DM}^2} S$.

Finally, for DM annihilations to $\chi_{1}^{\prime} \chi_{1}^{\prime}$, there is no guarantee of a light mediator, so a reasonable benchmark is $\langle \sigma v \rangle_{thermal}$, without any Sommerfeld enhancement.  However, it is possible that the present day cross section could differ from this benchmark.  For example, if the $m_Z^{\prime}$ is near $2 m_{\chi_1^{\prime}}$, the presence of this resonance can make the present day cross section either larger or smaller than $\langle \sigma v \rangle_{thermal}$, depending on whether the finite temperature in the early Universe pushes annihilations away from or closer to the resonance. 

\subsection{Analysis Details}\label{sec:analysis}

In this subsection, we provide details of our analysis to derive bounds on dark matter cross sections. 
The Poisson likelihood function for binned analysis is given by
\begin{equation}
\mathcal{L}(\mu \vert n) = \prod_{i,j} \frac{\mu_{ij}^{n_{ij}}}{n_{ij}!} \exp (-\mu_{ij}),
\label{eq:PoissonL}
\end{equation}
where $\mu_{ij}$ is the predicted number of events in the $i^{\rm th}$ energy bin and $j^{\rm th}$ region of interest, and $n_{ij}$ is the observed number of counts.  The prediction $\mu_{ij}$ is the sum of the counts calculated from background models as well as the dark matter signal with some annihilation cross section $\langle \sigma v \rangle$.  This likelihood function can be modified with additional factors to account for systematic uncertainties.  We define a test statistic,
\begin{equation}
    \label{eq:importantTestStatistic}
    TS = -2(\ln(\mathcal{L}(\mu \vert n))-\ln(\mathcal{L}(\hat{\mu} \vert n))).
\end{equation}
Here, $\hat{\mu}$ is the model prediction that maximizes the likelihood function.  95$\%$ confidence upper bounds can be placed on $\langle \sigma v \rangle$ by increasing $\langle \sigma v \rangle$ and hence $\mu$ from the best fit value until $TS = 2.71$. For placing projected bounds from CTA, the measured number of counts $n$ is simulated by an Asimov data set, which in this case is the mean number of counts expected from the background with no contribution from the dark matter signal.  Thus, the best fit value is $\hat{\mu} = n$, with $\langle \sigma v \rangle = 0$.  For our Fermi analysis, we maximize the likelihood with non-negative values of $\langle \sigma v \rangle$; this will place bounds that are at least as conservative as allowing an unphysical $\langle \sigma v \rangle < 0$ to be the best fit.

The Fermi bounds are derived based on 6 years of data from observations of 15 dwarf spheroidal (dSph) galaxies. Interpolation tables for the log-likelihood per energy bin are available from the Fermi-LAT Collaboration~\cite{Ackermann:2015zua}.  The interpolation tables are functions of the flux of energy over a single energy bin, $\int E \frac{d\Phi_{\gamma}}{dE} dE$, calculated assuming $\frac{dN}{dE} \propto E^{-2}$ over each energy bin.  For each dSph, we use the J-factor provided by the Collaboration in~\cite{Ackermann:2015zua}.  These are calculated assuming an NFW profile~\cite{Navarro:1996gj}, but the assumed profile should not make a significant difference. We sum over individual contributions from the 15 dSph to obtain the total log-likelihood for our TS in Eq.~(\ref{eq:importantTestStatistic}).

For the CTA projections, we assume a combined 525 hours of observation evenly distributed between 9 pointing positions near the Milky Way GC, centered on $l=\pm 1^{\circ}, 0^{\circ}$ and $b=\pm 1^{\circ}, 0^{\circ}$ in galactic coordinates.  We assume the GC DM density follows the Einasto profile given in Eq.~(\ref{eq:Einasto}).  Interpolation tables for the test statistic per energy bin are available from the CTA Consortium~\cite{Acharyya:2020sbj}.  The interpolation tables are a function of the flux of energy $E \frac{d\Phi_{\gamma}}{dE}$, evaluated at the geometric mean of the bounds of each energy bin, calculated assuming $\frac{dN}{dE} \propto E^{-2}$ over each energy bin. Backgrounds relevant in this analysis have been folded into the interpolation tables.

As a cross-check of the robustness of the above analysis, we constructed an alternate analysis framework based on an older projection for CTA bounds~\cite{Silverwood:2014yza}.  In our alternate analysis, we retain the statistical techniques of Ref.~\cite{Silverwood:2014yza}, but update specifications for CTA, background models, and search strategy.  We used the effective area and CR background for CTA-South from~\cite{CTAArea}, approximated astrophysical gamma ray backgrounds using the Fermi-LAT Collaboration's {\tt gll\_iem\_v07.fits} model~\cite{FermiBackground}, assumed 500 hours of observation time, and used regions of interest and energy bins similar to those of~\cite{Acharyya:2020sbj}.  Despite the simpler statistical and background treatment, our alternate analysis reproduced the bounds derived from the CTA Consortium's interpolation table to within $30\%$ for direct $b\bar{b}$ annihilation for $m_{DM} > 80$ GeV.

\subsection{Experimental Sensitivity}

We now present the current and projected experimental sensitivities to the above annihilation channels from Fermi-LAT and CTA. 

\begin{figure}\hspace{-0.35in}
    \begin{minipage}{0.5\textwidth}
        \vspace{0.19in}
        \includegraphics[width=1\textwidth]{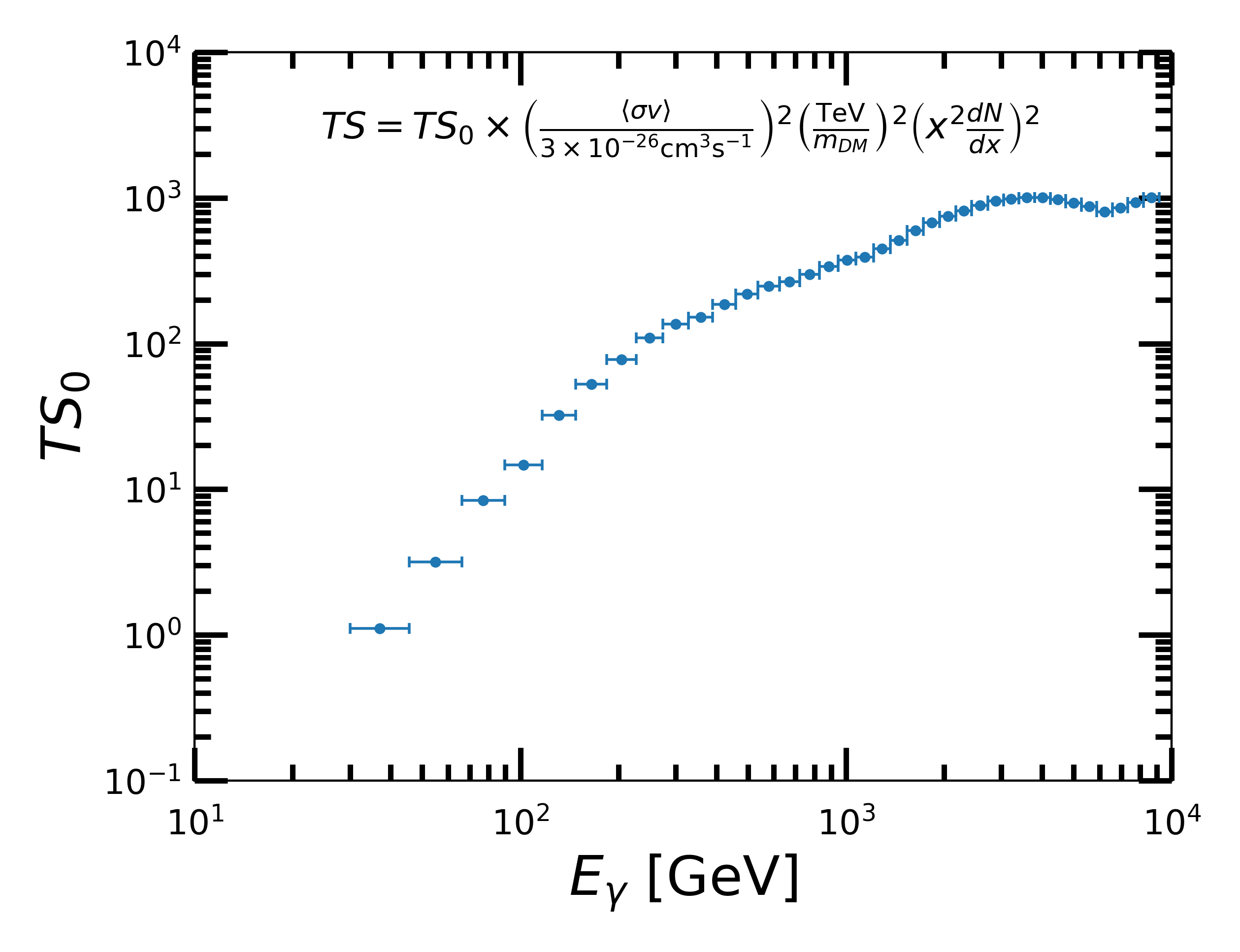}
%        \subcaption*{\textbf{a)}}
    \end{minipage}%
    \begin{minipage}{0.5\textwidth}
        \includegraphics[width=1.1\textwidth]{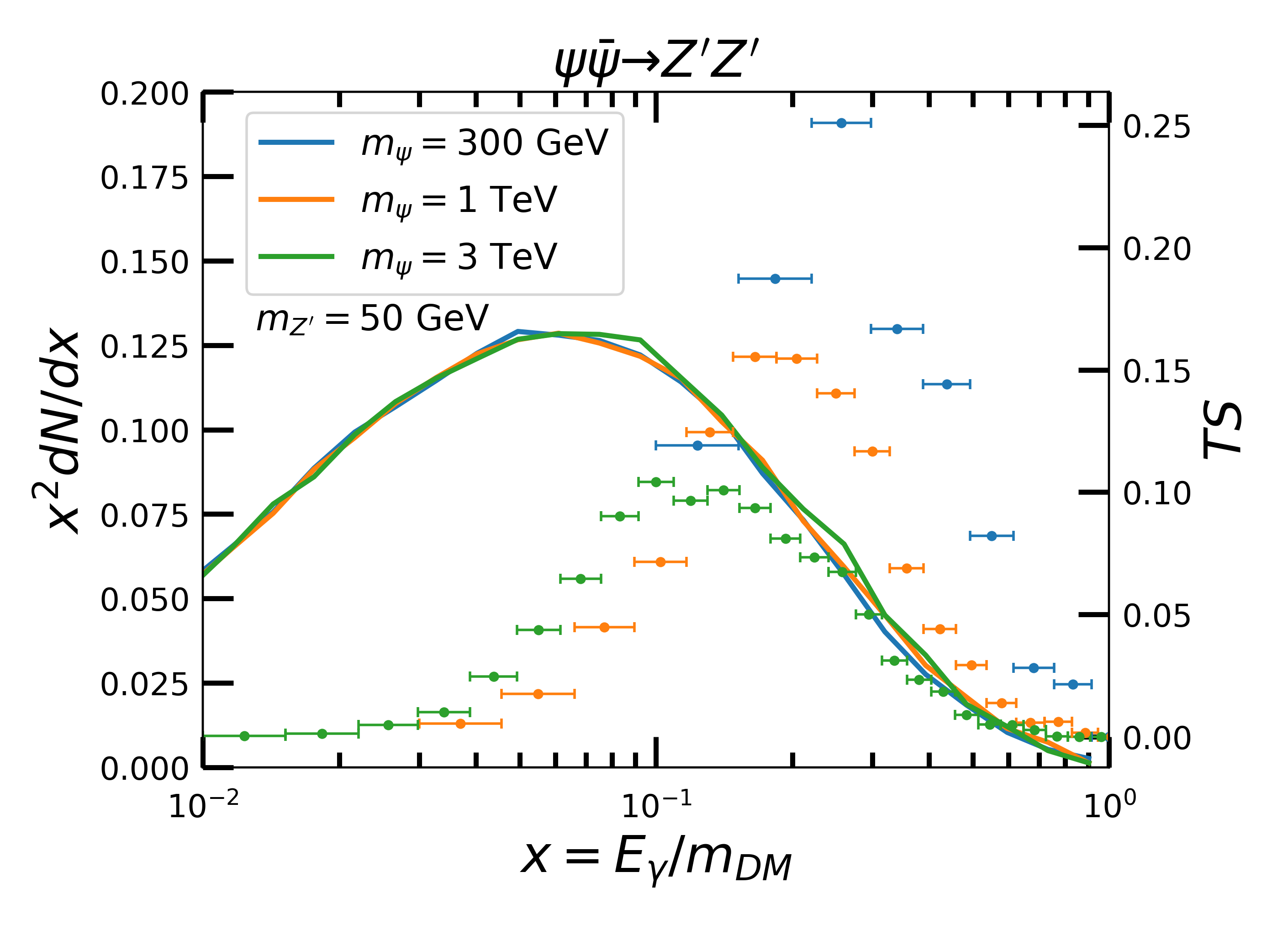}   
%        \subcaption*{\textbf{b)}}
    \end{minipage}
    \caption{ \textbf{Left:} Contributions to the test statistic, Eq.~(\ref{eq:importantTestStatistic}), of each energy bin for CTA assuming a flat photon spectrum, i.e., $x^2 dN/dx=1$. The ends of the bars denote the energy binning used in the interpolation tables. Using this flat spectra test statistic, $TS_0$, we can approximately (see text) recover the $TS$ for generic spectra as shown in the inset. \textbf{Right:} We show example spectra (solid curves) and their corresponding contributions to the test statistic (horizontal bars) for individual energy bins. We use cross sections that give the 95\% CL bound from CTA: $\langle\sigma v\rangle_{300 \text{ GeV}}=2.7 \times 10^{-26} \text{ cm}^3\text{s}^{-1}$, $\langle\sigma v\rangle_{1 \text{ TeV}}=1.8 \times 10^{-26}\text{ cm}^3\text{s}^{-1}$, and $\langle\sigma v\rangle_{3 \text{ TeV}}=2.0 \times 10^{-26}\text{ cm}^3\text{s}^{-1}$.}
    \label{fig:CTA Sensitivity new}
\end{figure}

For CTA, the effective area is a rapidly increasing function of energy. We illustrate the relative contributions of individual energy bins to projected CTA bounds in FIG.~\ref{fig:CTA Sensitivity new} (left panel) for a flat photon spectrum ($x^2 dN/dx=1$).  The y-axis is the contribution to the test statistic, normalized to show the dependence on the dark matter mass and cross section. 
Relative contributions of individual energy bins to the projected bounds for a few realistic DM spectra (solid curves) are shown in the right panel. In the limit where the number of background counts is much larger than the number of signal counts, and in the absence of correlated systematic errors between bins, the log-likelihood of each bin scales as the signal squared.  
One can extract approximate bounds for generic spectra using the inset text, as can be checked explicitly via comparison of the left and right panels. We find this scaling is only accurate to $\sim10\%$,  so our projected bounds are set using the full likelihood tables in~\cite{Acharyya:2020sbj}. These plots demonstrate that within the $100$ GeV - few TeV dark matter mass range, the projected reach of CTA is driven by the high energy tail rather than the peak of the gamma ray spectrum. This fact, and the sharp energy cutoff at $\sim 30$ GeV implies harder spectra will be probed more definitively by CTA.
If one constructed a plot similar to FIG.~\ref{fig:CTA Sensitivity new} (left panel) for Fermi-LAT, the contribution to the bound would peak around 2 GeV and then decrease at higher energies, unlike the case of CTA which increases before leveling off.  The contribution from the highest energy bin at Fermi, which reaches up to $500$ GeV, is only 3$\%$ that of the most important energy bin. 

\begin{figure}
    \includegraphics[width=0.7\textwidth]{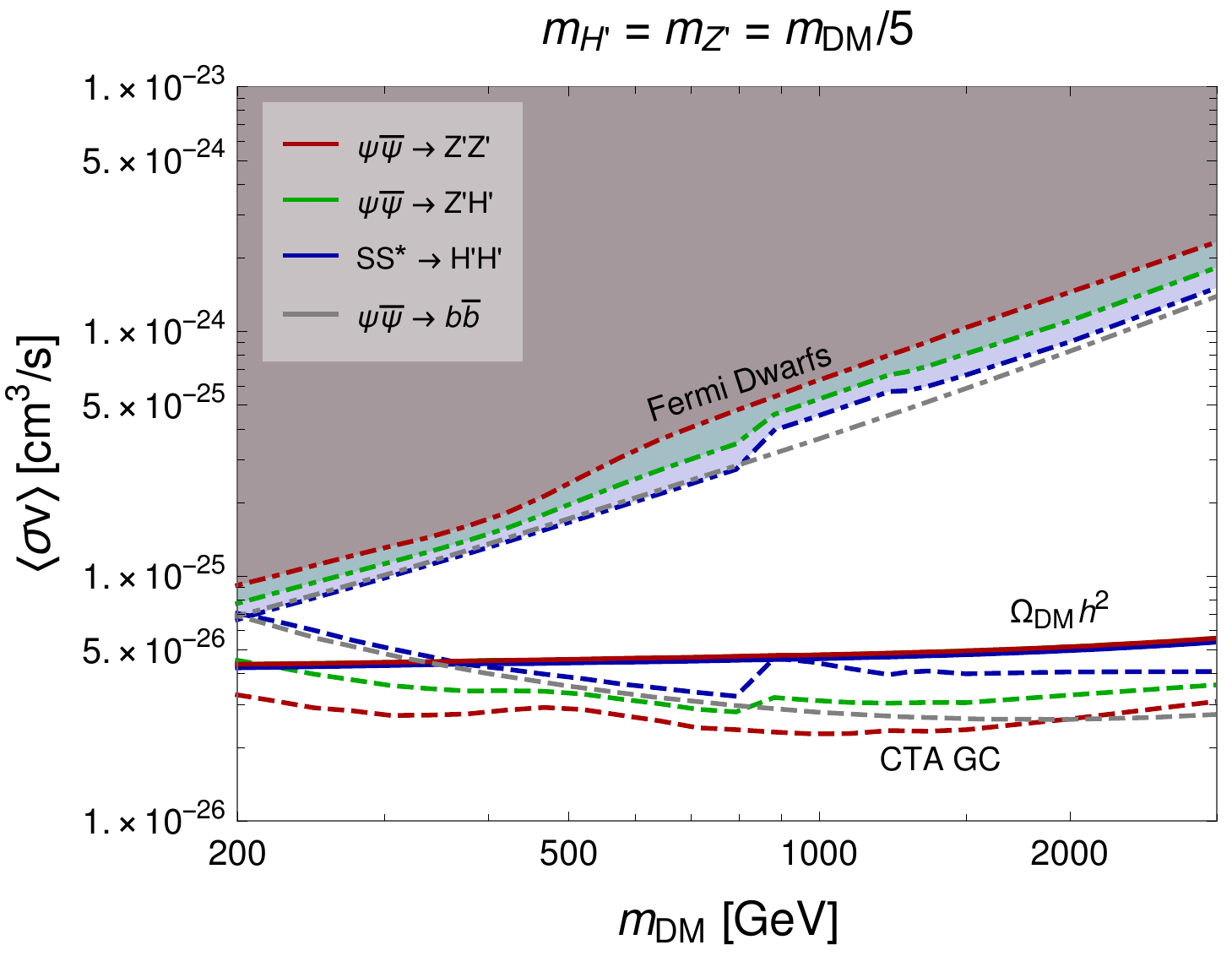}\\
    \vskip0.3cm
     \includegraphics[width=0.7\textwidth]{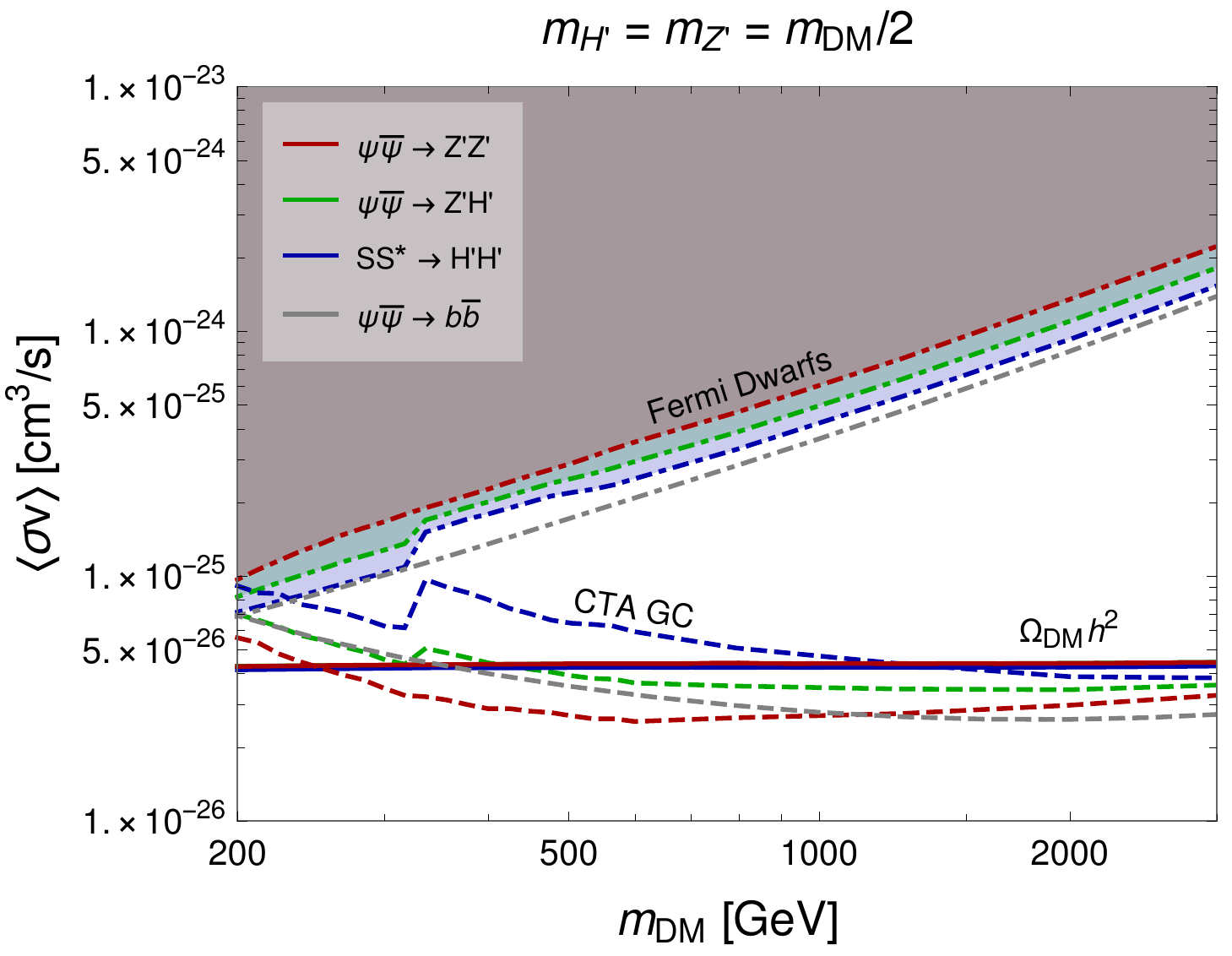}
    \caption{Current bounds from Fermi-LAT (dot-dashed curves) and projected CTA reach (dashed curves) for dark matter annihilation into secluded sector bosons, for $m_{DM}/m_\phi=5\, (2)$ in the top (bottom) panel.  Solid curves labelled  $\Omega_{DM} h^2$ show annihilation cross sections consistent with the correct thermal relic density. We also show limits/sensitivity curves for the benchmark case of direct annihilation to $b \bar{b}$ (grey).}
    \label{fig:ratiofixed}
\end{figure}

\begin{figure}[t!]
    \includegraphics[width=0.7\textwidth]{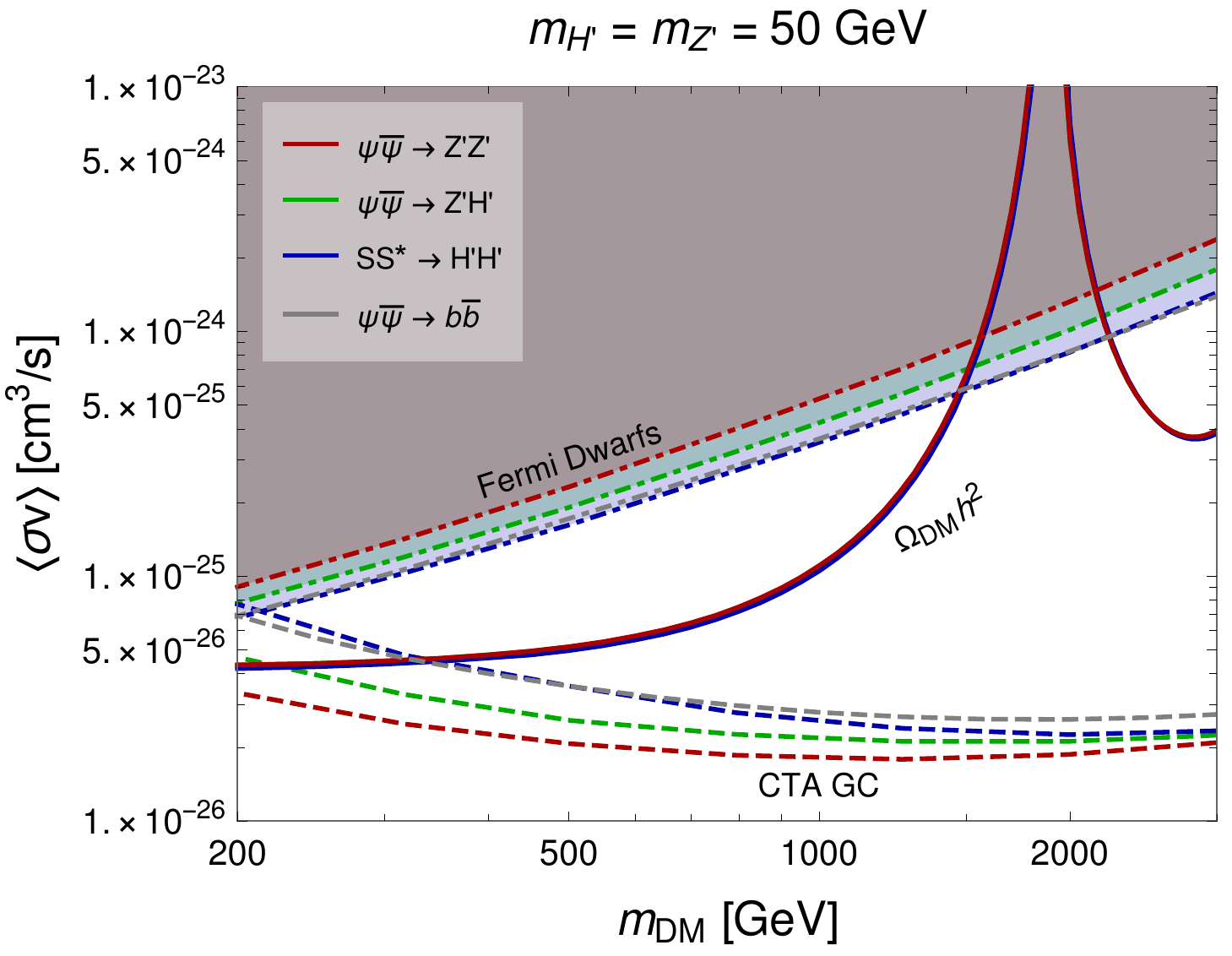}\\
    \vskip0.3cm
    \includegraphics[width=0.7\textwidth]{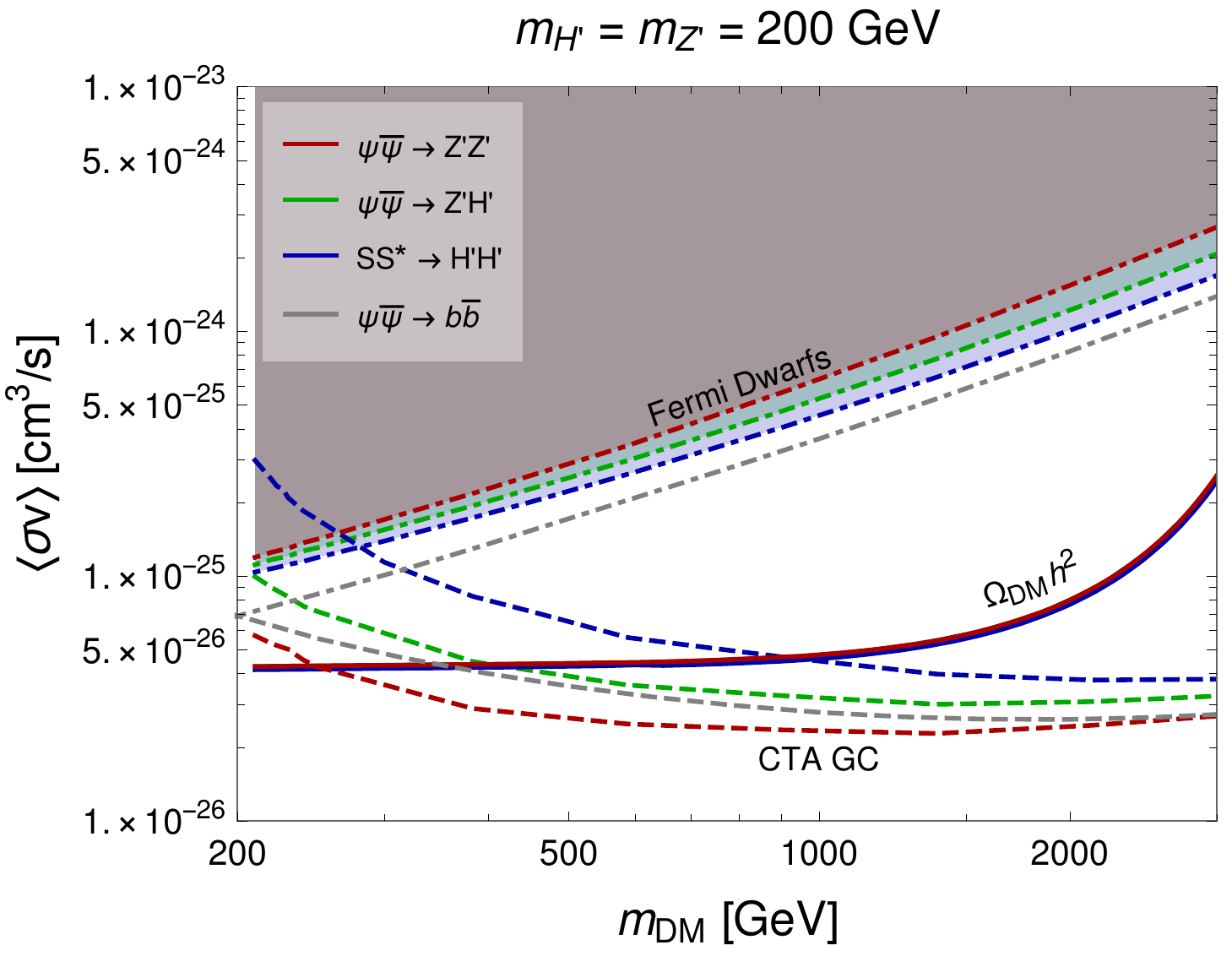}
    \caption{Same as FIG.~\ref{fig:ratiofixed}, but with fixed masses $m_\phi=50\,(200)$ GeV in the top (bottom) panel.  In the top panel, the $b\bar{b}$ Fermi Dwarfs curve is essentially coincident with the $H^{\prime} H^{\prime}$ curve. }
    \label{fig:massfixed}
\end{figure}

 We present the Fermi bounds and CTA projections for annihilations into secluded sector bosons ($H'H'$, $Z'Z'$, and $Z'H'$) in FIG.~\ref{fig:ratiofixed} (for fixed mass ratios $m_\phi/m_{DM}$; $\phi = H^{\prime}, Z^{\prime}$) and FIG.~\ref{fig:massfixed} (for fixed $m_\phi$).  
 In each case, the present day Sommerfeld-enhanced cross sections \textemdash ~corresponding to parameters that produce the correct thermal relic density, see Sec.~\ref{sec:sommerfeld} \textemdash ~are shown as solid curves and labelled as $\Omega_{DM} h^2$.  These curves correspond to Eqs.~(\ref{eq:ZH}), (\ref{eq:ZZ}), and (\ref{eq:H'H'AnnSpecial}), supplemented by the Sommerfeld enhancement of Eq.~(\ref{eq:Sommerfeld}). Sommerfeld enhancement is most pronounced for large ratios between the DM and mediator masses.  In FIG.~\ref{fig:ratiofixed}, the Sommerfeld effect is only discernible in the upper panel, and even then, only at the highest DM masses. Its effects are much more pronounced in FIG.~\ref{fig:massfixed}, particularly in the upper panel, because of the low mediator masses chosen. Note that the Sommerfeld enhancement depends on the dark matter velocity $v$ (see dependence on $\epsilon_v$ in Eq.~(\ref{eq:Sommerfeld}), which differs significantly between dwarf galaxies ($v\sim 10^{-4}$) and the galactic center ($v\sim 10^{-3}$)). However, the Sommerfeld enhancement factor (Eq.~(\ref{eq:Sommerfeld})) saturates to a constant value for $v\lesssim m_\phi/m_{DM}$, which is the case for all of our plots, so that the same $\Omega h^2$ cross section curve serves as a thermal DM target for both CTA observations of the galactic center and Fermi-LAT observation of dwarf galaxies.  The curves should be interpreted with caution near resonances, owing to the subtlety regarding kinetic decoupling discussed above.  In any case, we expect points near resonances to be observable with telescopes, independent of the details of kinetic decoupling. Finally, in all cases, we use the tree-level $s$-wave cross section; however, especially at large masses, where couplings are large, higher order effects are expected to be significant. 
 We see that Fermi does not constrain thermal relic cross sections except in narrow regions featuring resonant Sommerfeld enhancement. However, the thermal relic cross section is within the projected reach of CTA for all of these channels across a broad range of dark matter masses once the annihilation spectrum moves to sufficiently high energy, $m_{DM}\gtrsim$ few hundred GeV. 
 
The Fermi-LAT bounds are the most stringent for the $H'H'$ annihilation channel, followed by $Z'H'$ and $Z'Z'$.  The CTA constraints follow the opposite order.  This reversal can be understood in terms of the spectra produced by the annihilation products, coupled with an understanding of which parts of the spectra each experiment is sensitive to.  
As shown in FIG.~\ref{fig:CTA Sensitivity new}, CTA is more sensitive to the higher energy tail of the spectrum than the peak.  Fermi, on the other hand, is already quite sensitive to $\sim$ GeV photons, hence its limits depend more on the overall photon count.
FIG.~\ref{fig:ZH_ZZ} shows that final states with $H'$ tend to produce spectra with higher peaks, while final states with $Z'$ tend to create harder tails. Therefore, the projected CTA reach is strongest for the $Z'Z'$ channel, probing thermal relics for $m_{DM}\gtrsim 200$ GeV, and grows weaker for $Z'H'$ and further for $H'H'$, for which sensitivity to thermal relics is only achieved for masses closer to the TeV scale. 

In FIG.~\ref{fig:ratiofixed}, the $S S^* \rightarrow H'H'$ and $\psi \bar{\psi} \rightarrow Z'H'$ curves have features around $m_{DM} = 800~(320)$ GeV in the top (bottom) panel, corresponding to the $H' \rightarrow WW$ decay channel becoming kinematically accessible.    The $H'$ inherits the decay channels of the SM Higgs boson $h$ through the portal mixing, so the $H'\to b \bar{b}$ channel dominates below this kinematic threshold, whereas $H^{\prime} \rightarrow WW$ dominates above the threshold, worsening the reach by an $\mathcal{O}(1)$ factor (this is also clear from comparing the CTA curves in the two panels in FIG.~\ref{fig:massfixed}). Since $H'\to WW$ remains the dominant channel, crossing subsequent kinematic thresholds for $H^{\prime} \to ZZ, hh, t\bar{t}$ do not produce noticeable effects on the experimental sensitivities. 

As the $Z'$ decays primarily to up-type quarks and charged leptons at most masses (recall that it couples to electric charge (hypercharge) in the limit $m_{Z'}\ll m_Z$ ($m_{Z'} \gg m_Z$)), there are no major features in the $\psi \bar{\psi} \rightarrow Z'Z'$ bound plot.  However, for $m_{Z'} \simeq m_{Z}$, down quarks dominate and neutrinos can be important.  This leads to a modest feature on the $Z'Z'$ curve in the top panel of FIG.~\ref{fig:ratiofixed} near $m_{DM}\approx 460$ GeV.

\begin{figure}[tp]
%    \centering
    \includegraphics[width=0.7\textwidth]{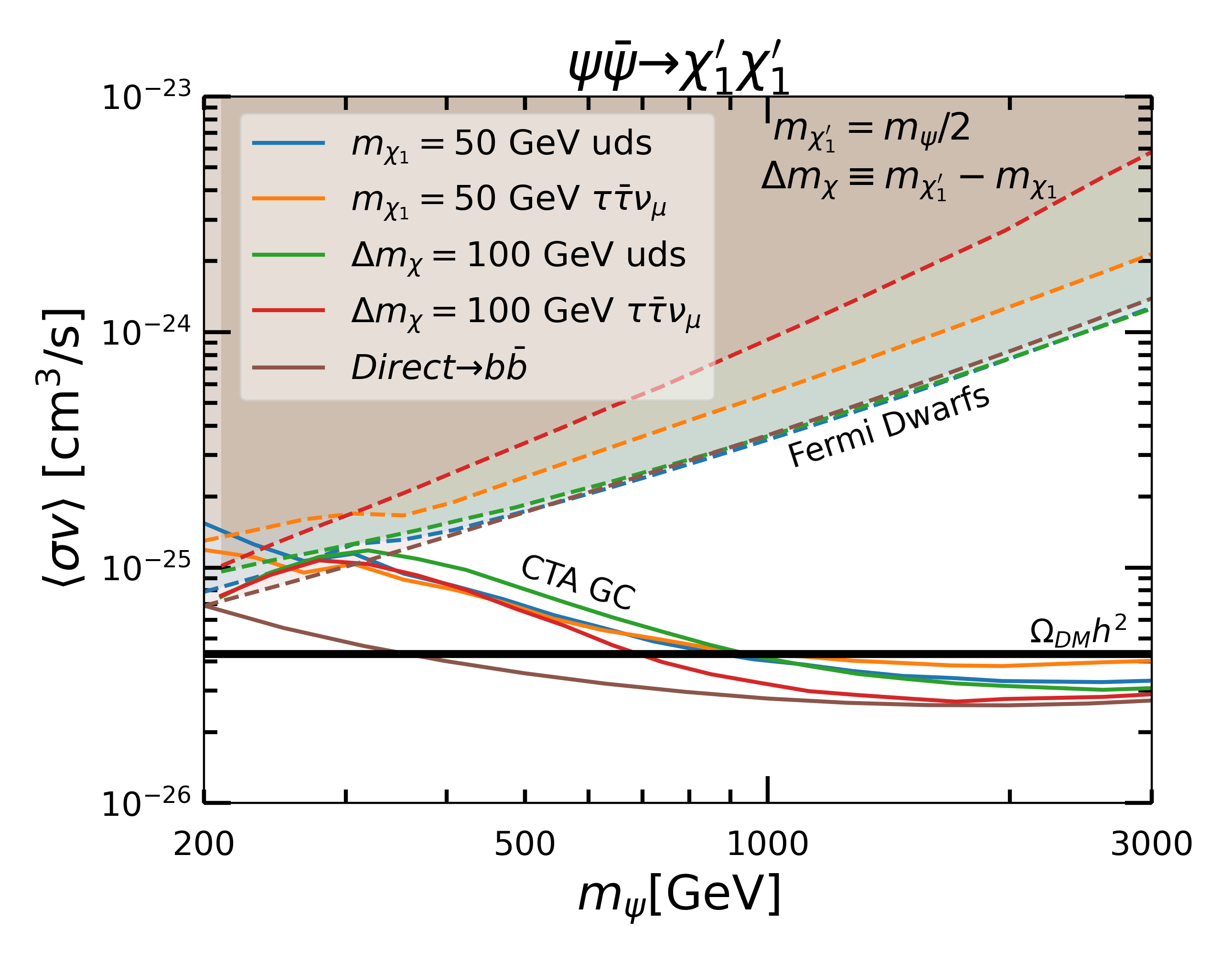}\vspace{-0.15in}
    \includegraphics[width=0.7\textwidth]{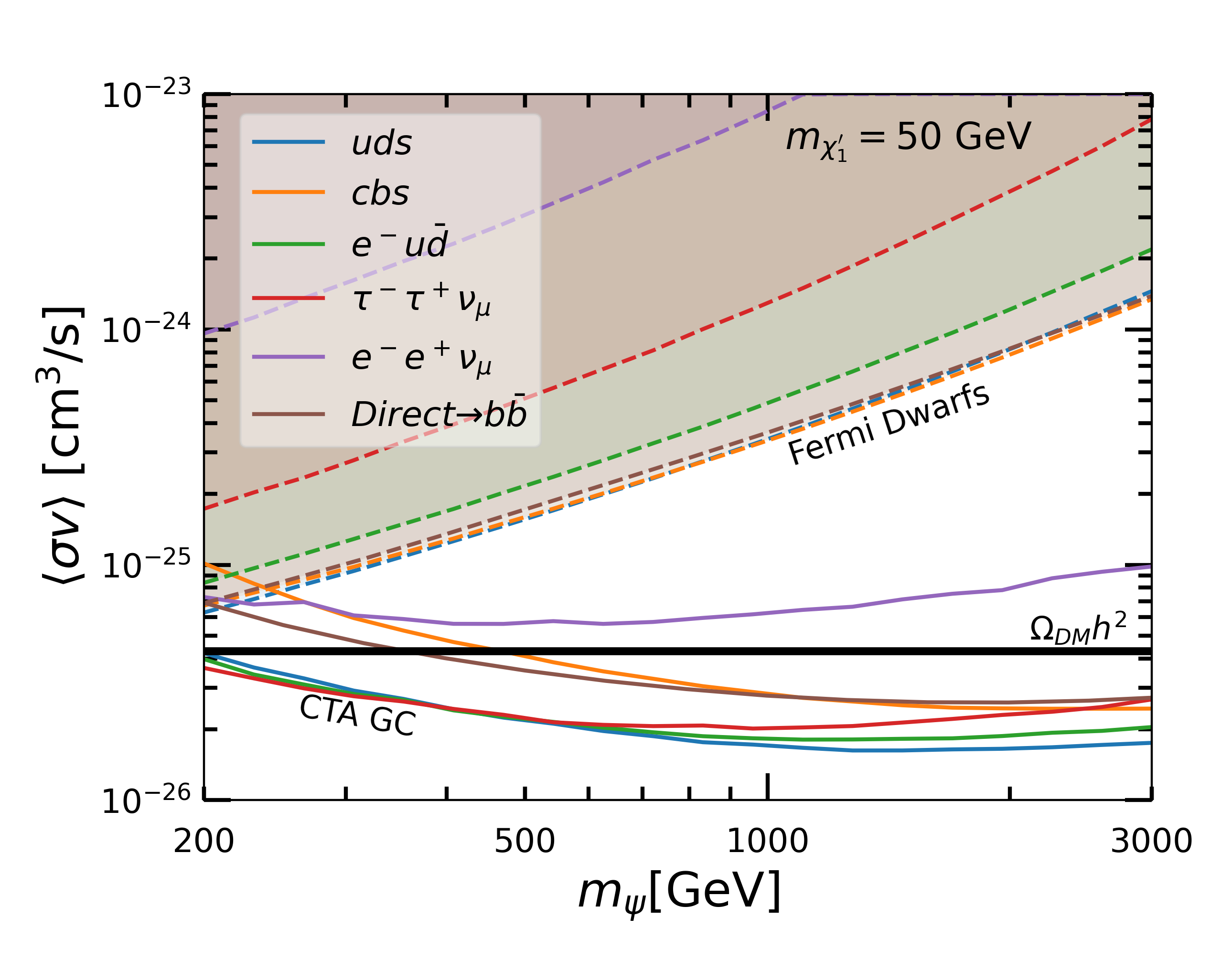}
    \caption{Fermi Dwarf limits (dashed) and CTA GC (solid) sensitivity projections for annihilation to secluded neutralino states. Also shown: the thermal cross section (solid black, labelled $\Omega_{DM}h^2$), and limits/sensitivity curves for the benchmark case of direct annihilation to $b \bar{b}$ (brown). \textbf{Top:} 
    $\chi_1^\prime$ decays to LSP $\chi_1$ and either $Z^{(\ast)}$, $h^{(\ast)}$. The $\chi_1$ subsequently decays via RPV coupling $\lambda''_{112}$ or $\lambda_{313}$. Mass spectrum and branching ratio information is in the text. \textbf{Bottom:} The LSP is $\chi_1^\prime$, which decays directly to the final state specified in the legend via RPV. }
    \label{fig:Bounds all chi}
\end{figure}

In FIG.~\ref{fig:Bounds all chi}, we show analogous results for annihilation into secluded sector neutralinos, $\psi\bar{\psi}\to \chi_1'\chi_1'$, for cases where $\chi_1'$ decays to $\chi_1$, which subsequently decays via RPV couplings (top panel), and where $\chi_1'$ is the LSP and decays directly via RPV couplings (bottom panel). For reference, we also show the reach for direct annihilation into $b\bar{b}$ (brown curves) as well as the thermal cross section (solid black line). These plots show several similarities with the analogous plots for annihilation into secluded sector bosons in FIG.~\ref{fig:ratiofixed},\,\ref{fig:massfixed}: we see that the relative ordering of sensitivity to various channels differs between Fermi and CTA due to the two instruments being sensitive to different parts of the produced gamma ray spectra; and the thermal cross section, while out of reach of Fermi, can be probed by CTA for $m_{DM}\gtrsim$ few hundred GeV. 

In the top panel, which features $\chi_1'\to \chi_1 (Z/h)$ decays with subsequent $\chi_1$ decay via RPV couplings,
limits for two sets of mass spectra are shown: one with fixed $m_{\chi_1}=50$ GeV (blue, orange), the other with $\Delta m_{\chi} \equiv m_{\chi'_1}-m_{\chi_1}=100$ GeV (green, red). For the latter curves, the mass splitting fixes $BR(\chi'_1 \to \chi_1 Z)=100\%$.  We can see several interesting features in this plot. The CTA sensitivity to the $uds$ RPV decay channel improves relative to that for the $\tau\tau\nu$ channel as the dark matter mass increases. The gamma ray spectrum for $uds$ possesses a higher peak but fewer high energy photons than the $\tau\tau\nu$ spectrum. Increasing the DM mass makes CTA, with its relatively high energy threshold, increasingly sensitive to the peak. Likewise, for the leptonic channel, we see that CTA has improved sensitivity when $\chi_1$ is heavier as the gamma ray spectrum is harder in this case (see FIG.~\ref{fig:RPV Mass vary}, left panel), whereas it is relatively insensitive to such variations for the $uds$ case (see FIG.~\ref{fig:RPV Mass vary}, right panel).  In the case where the $m_{\chi_1}$ is fixed (blue, orange), the branching ratios for $\chi_1^{\prime}$ decay will change as $m_{\psi}$ varies.  For $m_\psi < 280$ GeV, $\chi'_1< 140$ GeV decays via 3-body processes to SM fermion pairs and $\chi_1$. This is reflected in the kink in these curves at $m_{\psi} \simeq 280$ GeV.  For higher masses, up until the $h$ is on-shell at $m_\psi=350$ GeV, the secluded neutralino decays exclusively to $\chi_1 Z$. For even heavier $\chi'_1$, $BR(\chi'_1\to \chi_1 h)$ rapidly approaches $85\%$, then slowly declines to $64\%$ at the right edge, with the rest of the branching to $\chi_1 Z$. 
Overall, CTA sensitivity is weaker for cases where $\chi_1^{\prime}$ decays to an MSSM LSP than for DM direct annihilation into $b\bar{b}$, as the multiple final states from the cascades result in softer spectra.  Nevertheless, CTA is able to probe thermal cross sections for $m_{DM}\gtrsim 700$ GeV.  

\begin{figure}[tp]
%    \centering
    \includegraphics[width=0.8\textwidth]{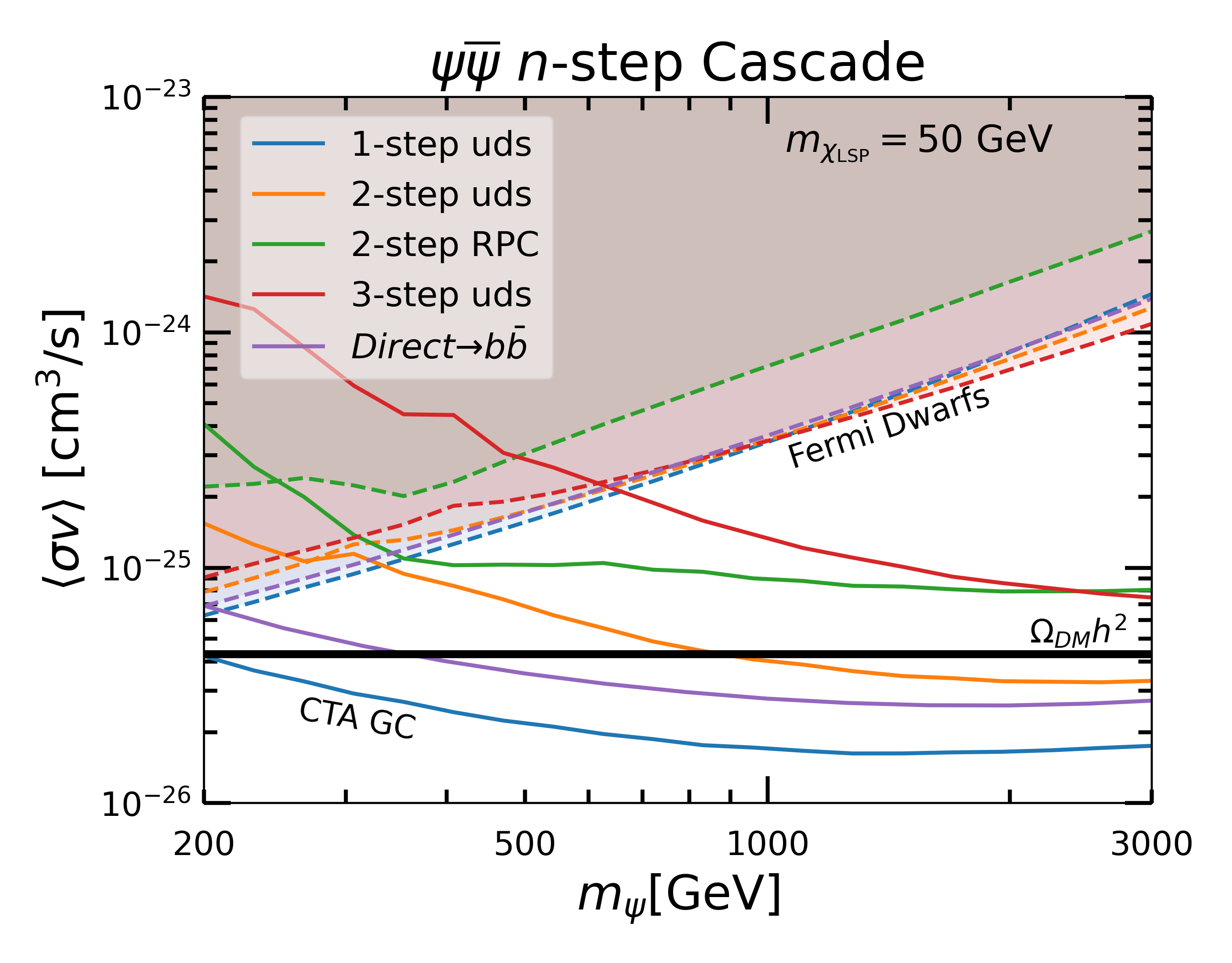}
    \caption{Fermi Dwarf limits (dashed) and CTA GC (solid) sensitivity projections for annihilation to secluded neutralino states. Also shown: the thermal cross section (solid black, labelled $\Omega_{DM}h^2$), and limit/sensitivity curves for the benchmark case of direct annihilation to $b \bar{b}$ (purple). The $n$-step notation is as in FIG.~\ref{fig:Photon Step Comparison}; see that caption for details of the cascades. We generalize those masses to $m_{\chi'_1}=m_\psi/2$ for the 2-step case, and $m_{\chi'_1}=3 m_\psi/8$, $m_{Z'}=7m_\psi/8$ for the 3-step case. The branching ratios of $\chi'_1$ decay in these cases are identical to those in FIG.~\ref{fig:Bounds all chi}.   Here $m_{LSP}$ can refer to either the $\chi_{1}$ (2, 3-step) or the $\chi_1^{\prime}$ (1-step) mass.}
    \label{fig:Bounds cascade chi }
\end{figure}

Next, we discuss the results in the bottom panel, where $\chi_1'$ decays directly via RPV couplings. Similar to the top panel, we see CTA has progressively better sensitivity to $uds$ compared to $\tau\tau\nu$ coupling for heavier dark matter. CTA is also more sensitive to spectra with lighter $m_{\chi_1'}$, as the decay products are more boosted. 
It is interesting to note that, in contrast to the $\chi_1$ LSP case, for this $\chi_1'$ LSP scenario CTA sensitivity can be even $\textit{better}$ than for direct annihilation into $b\bar{b}$ for the $uds$ and $\tau\tau\nu$ RPV couplings, which produce pions copiously that can be highly boosted. Indeed, for these couplings, CTA can probe the thermal relic cross section for the entire mass range we consider. However, for other choices of RPV couplings that do not (directly) produce pions, such as $cbs$ and $ee\nu$, sensitivity can significantly weaken: as we see from the plot, CTA is unable to probe $ee\nu$ decays in any part of the parameter space we consider.    

 We also place projected constraints on cases where R-parity is conserved, and where an additional cascade step is present, shown in FIG.~\ref{fig:Bounds cascade chi }. This plot utilizes spectra analogous to those shown previously in FIG.~\ref{fig:Photon Step Comparison}.  In particular, that figure shows precisely the spectra used to extract bounds at $m_{DM}=3$ TeV. The spectra change only at the few percent level for  $m_\psi \gtrsim 1000$ GeV, thus the changes in the bounds in this regime are largely from the energy dependence of CTA's sensitivity. Mass thresholds and kinematics are significant for $m_\psi \lesssim 500$ GeV in the 2 and 3 step cases; however, this regime is likely unobservable at CTA for such topologies. We see that decays with fewer steps have stronger projected constraints over the mass range we consider, due to the rapid increase in sensitivity of CTA for photons with energies $\gtrsim$ 100 GeV. For the spectrum chosen here, the RPC case is unobservable for any DM mass.  However, while these spectra are representative benchmark points in parameter space, in small regions of parameter space even RPC cases could have observable thermal cross sections at CTA. The accessibility of such regions would hinge upon the fraction of the DM comprised of the RPC LSP.
Note that the RPC case, while much more difficult to bound than its RPV counterpart due to the LSP being invisible (compare green and orange curves),  can be potentially more observable than the RPV case with an additional cascade step (compare green and red curves), simply because more high energy photons can be present in cascades with fewer steps.

\section{Summary} \label{sec:summary}
If dark matter resides in a secluded sector, direct detection signals will be suppressed by potentially tiny portal couplings between the secluded and visible sectors, but thermal annihilation cross sections relevant for indirect detection can remain comparable to visible sector WIMP models.  Novel annihilation spectra are possible, which can impact the sensitivity of current and upcoming experiments to these models. 
The spectra and the resultant limits depend on the details of the model.  Supersymmetric models give a particularly attractive realization of secluded sector WIMP dark matter matter, as supersymmetry breaking can explain the closeness of the scale in the secluded sector to the weak scale.  In this case, matter in the secluded sector can supplant the LSP as a dark matter candidate.  This provides renewed motivation for models of $R$-parity violating supersymmetry.  In cascades that include the LSP, the dominant RPV coupling impacts the final photon spectra from dark matter annihilation.  

In this paper, we have explored a variety of these cascade spectra, and derived limits from Fermi dwarf galaxy observations and projections for CTA observations of the galactic center. We find that CTA should be sensitive to a wide range of models of this type, with the precise reach depending on the details of the mass spectra and the branching ratios of the annihilation products. For models where the DM annihilates to secluded bosons that subsequently decay via portals into SM states, dark matter masses of several hundred GeV and upwards can be probed as long as the secluded bosons are not too heavy.  When DM annihilation spectra depend on $R$-parity violating couplings, the reach can vary dramatically.  Cases with relatively hard spectra \textemdash ~for example, those involving $\tau$ leptons or light quarks \textemdash ~can be probed over a wide mass range.  Indeed, we find that CTA can be more sensitive to these scenarios than other popular benchmarks like direct annihilations to $b$ quarks.  Those involving fewer hard photons, for example, annihilations that involve electrons, can be much more challenging to probe. We emphasize that our results, although derived within a supersymmetric framework, are more generically applicable to a broader variety of secluded dark matter scenarios with cascade decays.

We have ignored the possibility that secluded sector particles decay to any MSSM particles except the LSP.  If such particles were kinematically accessible, this could lead to longer decay chains.  It might be of interest to study this case in more detail.  As we have shown, longer decay chains lead to softer spectra, so we might expect them to be more difficult to probe with CTA. But there is a compensating effect: decays to other supersymmetric particles are most likely for heavy dark matter, which would result in more energetic photons, to which CTA is more sensitive.   

It would be of interest to consider signals arising from other final states in these setups, including anti-protons and positrons, as well as limits arising from the impact of dark matter annihilations on the cosmic microwave background. Collider searches for RPV decays of visible sector particles could also provide complementary probes of such frameworks; while such a discovery carries no obvious connection to dark matter, the discovery of RPV SUSY would necessitate the existence of additional particles and symmetries beyond the MSSM to explain dark matter and would be strongly suggestive of the kind of model studied here.

\section*{Acknowledgments} 
%%%%%%%%%%%%%%%%%%%%%%%%%%%%
We thank J.~Corsello for collaboration in early stages of this work.  The work of PB, ZJ, and AP is supported in part by the DoE grant DE-SC0007859. AP would also like to thank the Simons Foundation for support during his sabbatical. The work of BS is supported by the Deutsche Forschungsgemeinschaft under Germany’s Excellence Strategy - EXC 2121 Quantum Universe - 390833306.

\bibliography{bib}

\end{document}